\documentclass{article}
\usepackage[utf8]{inputenc}
\usepackage{verbatim}
\usepackage{listings}
\usepackage[letterpaper, margin=1in]{geometry}
\usepackage{braket}
\usepackage{enumitem, kantlipsum}
\usepackage{graphicx}
\usepackage{caption}
\usepackage{subcaption}
\usepackage[dvipsnames,table,xcdraw]{xcolor}
\usepackage{multicol}
\usepackage{mathtools}

\usepackage{csquotes}
\usepackage{amsmath}

\usepackage{soul}
\usepackage{url}

\definecolor{codegreen}{rgb}{0,0.6,0}
\definecolor{codegray}{rgb}{0.5,0.5,0.5}
\definecolor{codepurple}{rgb}{0.58,0,0.82}
\definecolor{backcolour}{rgb}{0.95,0.95,0.92}
\definecolor{codeblue}{rgb}{0.92,0.94,0.97}
\definecolor{light-gray}{gray}{0.95}

\lstdefinestyle{mystyle}{
    backgroundcolor=\color{backcolour},   
    commentstyle=\color{codegreen},
    keywordstyle=\color{magenta},
    numberstyle=\tiny\color{codegray},
    stringstyle=\color{codepurple},
    basicstyle=\footnotesize,
    breakatwhitespace=false,         
    breaklines=true,                 
    captionpos=b,                    
    keepspaces=true,                 
    numbers=left,                    
    numbersep=5pt,                  
    showspaces=false,                
    showstringspaces=false,
    showtabs=false,                  
    tabsize=4
}
 
\title{QuASeR\\Quantum Accelerated De Novo DNA Sequence Reconstruction}
\author{Aritra Sarkar \ \ \ \ Zaid Al-Ars \ \ \ \ Koen Bertels\\
Department of Quantum \& Computer Engineering\\
Delft University of Technology, The Netherlands}
\date{\today}

\begin{document}

\sloppy

\maketitle

\begin{abstract}
    In this article, we present QuASeR, a reference-free DNA sequence reconstruction implementation via de novo assembly on both gate-based and quantum annealing platforms.
    Each one of the four steps of the implementation (TSP, QUBO, Hamiltonians and QAOA) is explained with simple proof-of-concept examples to target both the genomics research community and quantum application developers in a self-contained manner.
    The details of the implementation are discussed for the various layers of the quantum full-stack accelerator design.
    We also highlight the limitations of current classical simulation and available quantum hardware systems.
    The implementation is open-source and can be found on \url{https://github.com/prince-ph0en1x/QuASeR}.
\end{abstract}


\section{Introduction} \label{s1}

Understanding the genome of an organism reveals insights with scientific and clinical significance like causes that drive cancer progression, intra-genomic processes influencing evolution, enhancing food quality and quantity from plants and animals.
Genomics data is projected to become the largest producer of big data within the decade~\cite{stephens2015big}, eclipsing all other sources of information generation, including astronomical as well as social data.
At the same time, genomics is expected to become an integral part of our daily life, providing insight and control over many of the processes taking place within our bodies and in our environment.
An exciting prospect is personalized medicine~\cite{hamburg2010path}, in which accurate diagnostics can identify patients who can benefit from precisely targeted therapies.
Despite the continual development of tools to process genomic data, current approaches are yet to meet the requirements for large-scale clinical genomics.
In this case, patient turnaround time, ease-of-use, robustness and running costs are critical.
As the cost of whole-genome sequencing (WGS) continues to drop~\cite{wgscost2018}, more and more data is churned out creating a staggering computational demand.
Therefore, efficient and cost-effective computational solutions are necessary to allow society to benefit from the potential positive impact of genomics.
This paper provides more efficient solutions to the high computational demands in the field of genomics.

\begin{figure*}[ht]
    \centering
    \includegraphics[clip, trim=0cm 3cm 0cm 3cm,  width=0.85\textwidth]{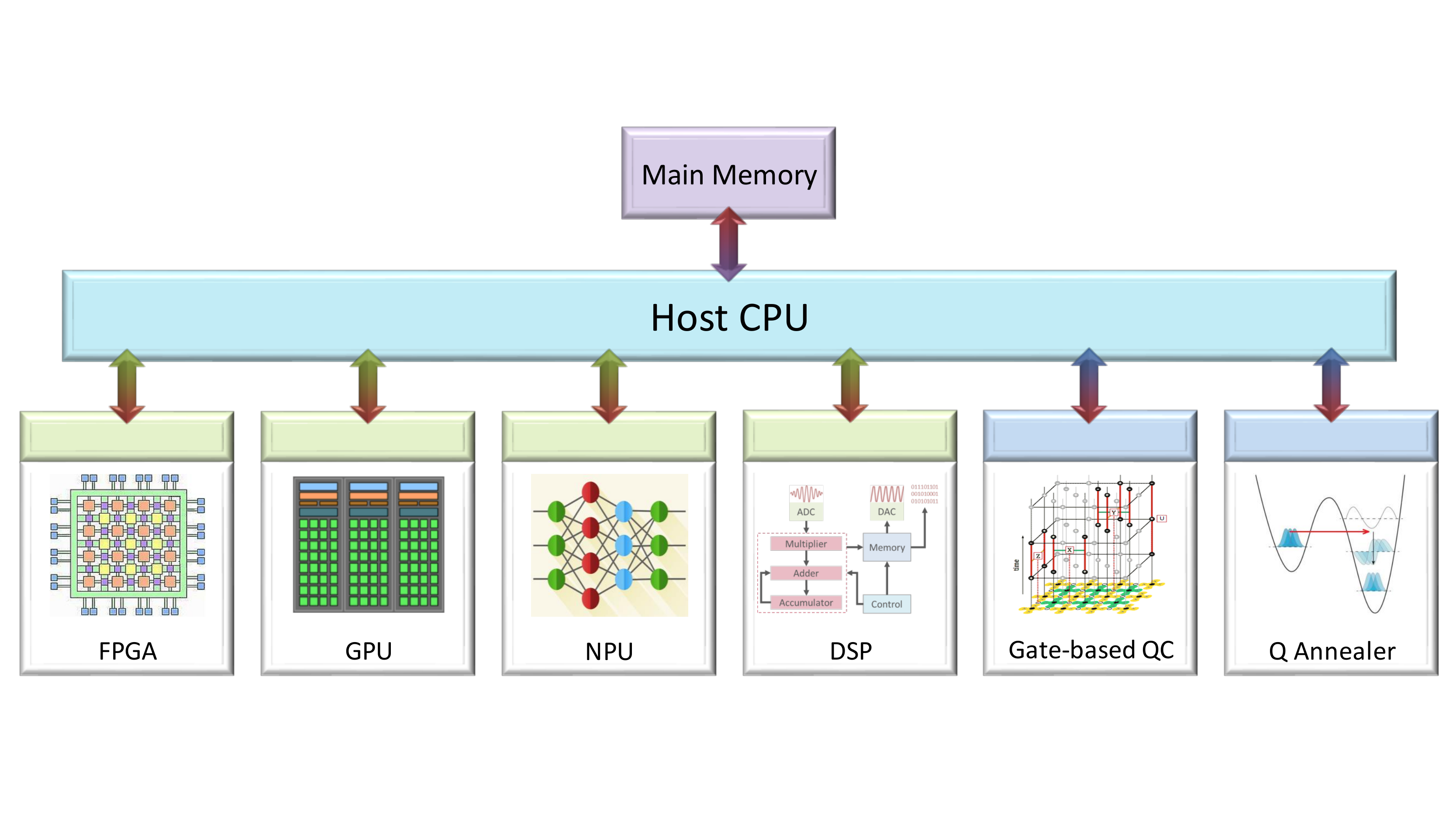}
    \caption{The accelerator model of computing}
    \label{f_accel}
\end{figure*}

The last decade in computer architecture has focused on the emergence of accelerators~\cite{vassiliadis2004} as specialized processing units to which the host processor offloads suitable computational tasks.
This is motivated by the various technological bottleneck (such as power use, frequency, memory access, level of parallelism) to increasing the computational capabilities of generic CPUs. 
Different accelerators, as shown in figure~\ref{f_accel}, are chosen based on the their strengths that enable better execution of a particular type of logical manipulation, as to speed up the overall execution time according to Amdahl's law.
Commonly used accelerators today include field-programmable gate arrays~(FPGA), graphics-processing units~(GPU), neural processing units~(NPU), digital signal processors~(DSP), etc.

An alternate computing paradigm that is receiving a lot of attention lately in computer architecture is quantum computing - which uses fundamental properties of quantum mechanics to achieve a computational advantage over classical computation.
Within the next decade, it is likely that applications will be a hybrid combination of a classical computer and a quantum accelerator, with multiple computational kernels, than a universal quantum processing unit.
As a holistic view we consider two classes of quantum accelerator as additional co-processors.
One is based on quantum gates and the second is based on quantum annealing.
The classical host processor keeps the control over the total system and delegates the execution of certain parts to the available accelerators.

In this paper, we propose an implementation of a DNA assembly technique (called de novo assembly) on a quantum computing platform.
This sequencing technique has many advantages over other simpler methods, but suffers from large computational complexity, which motivates targeting a quantum accelerator. 
We target both a gate-based quantum system as well as a quantum annealer.
Each step of the formulation is explained with simple examples to target both the genomics research community and quantum application developers.
The implementation is evaluated on the D-Wave simulator, the D-Wave annealer in the cloud and the QX Simulator.
The current limitations in solving real problem sizes to achieve a quantum advantage are discussed.
It is the first time this important computational problem in bioinformatics is targeted on a quantum accelerator.

This paper is organized as follows.
In section~\ref{s2}, we introduce the specific problem of sequence reconstruction via de novo assembly using the overlap-layout-consensus approach.
This is the target algorithm for which the quantum kernel is formulated.
In section~\ref{s3}, the quantum full-stack layers are presented to highlight how the algorithm logic interacts with the underlying classical and quantum system.
Section~\ref{s4} first introduces the required technical background for the tools for formulating the de novo assembly problem, i.e. QUBO, TSP and Hamiltonians.
A simple example for mapping a QUBO to the D-Wave quantum annealer is shown.
The variational algorithm approach for gate-based quantum computing is introduced for solving optimization problems using QAOA.
In section~\ref{s5}, the de novo assembly problem is mapped to a TSP and then the TSP is mapped to a QUBO.
A small proof-of-concept example is detailed at each step.
Section~\ref{s6} systematically solves the QUBO formulation first on the D-Wave simulator and annealer system, and thereafter a QAOA on OpenQL is attempted.
Then we discuss the limitations for scaling the formulation on real datasets using classical simulation and available quantum hardware systems.
Section~\ref{s7} concludes the paper. 
\section{Sequence reconstruction} \label{s2}

In order to sequence the DNA of an organism, DNA is broken down into fragments since DNA sequencing machines are not capable of reading the entire genome at once.
Then these fragments are sequenced using modern sequencing technologies (such as Illumina), which produces reads of approximately 50-150 base pairs at a time, with some known error rate.
To reconstruct the DNA from these fragments for further analysis, two different techniques are used (a) de novo reference-free assembly of reads and (b) ab initio reference-based alignment of reads.
Since the principles of quantum computation are fundamentally different, we will investigate the most basic algorithmic primitive for which the quantum kernel can be constructed.
Thus, before presenting the corresponding quantum algorithm, the existing classical algorithms are reviewed here.

\subsection{Ab initio reference-based alignment}

In the ab initio method, the DNA reads are matched to a trusted existing reference of the organism.
This is similar to a pattern matching problem, of finding the index of the read in the reference.
However, this method introduces bias based on the reference since, ironically, the next step after reconstruction is to discover variation from the reference for identifying implications.
Even the ab initio method is computationally infeasible for exact matches and thus heuristic methods are employed in state of the art tool-chains.

In the naive approach to sub-string matching involves matching the (relatively short) pattern string in a reference string.
The pattern is placed at the first index, and each consecutive characters in the pattern and the corresponding character in the reference is compared.
If the end of the pattern is reached, the index is returned with a success flag, else, the pattern is shifted by one place and the comparison is restarted from the beginning of the pattern.
Quantum variants of reference-based alignments is explored further in our work in~\cite{sarkar2019algorithm}.

The core idea of improving on the naive approach is developing a strategy such that, after a mismatch, more than one place shift is done to the pattern comparison position.
Performances of exact string matching algorithm like Boyer-Moore and Knuth-Pratt-Morris~\cite{gusfield1997algorithms} degrades quickly as they are not suitable for an approximate match which are very common in DNA sequences due to read errors.
The minimum alignment cost for Global Sequence Alignment can be found by the Needleman-Wunsch algorithm or the Local Sequence Alignment maximized over all possible substrings by the Smith-Waterman algorithm.
The industrial approaches (like BWA-MEM) are heuristics built on these basic algorithms to reduce cost in terms of memory and processing trading accuracy.

There are many different algorithms for alignment with no single winner as different heuristics work for different sequences and applications.
Current techniques trade speed and memory for accuracy.
These heuristics make the problem tractable for higher sized genomes like that of a human.
However, the approximations and errors introduced prevent further progress in critical application domains like personalized medicine.
Given enough computing power, de novo sequencing is highly desirable.

\subsection{De novo reference-free assembly}

Normally multiple copies of the DNA are made before fragmenting it.
Thus, a portion of the data will be preserved in multiple copies which are chopped off at different places resulting in data overlaps which facilitate stitching.
This method is called de novo reconstruction, as no other data than the sequenced read is used for reconstruction.
It is computationally expensive and done normally for the first time a new species is sequenced.

There are different methods~\cite{khan2018comprehensive} for de novo sequencing used by the available tools: Overlap-Layout-Consensus (OLC) methods, de Bruijn graph (DBG) methods, string graphs, greedy and hybrid algorithm, etc.
Real-world WGS data induces various problems in all these methods.
Examples are spurs (short, dead-end divergences from the main path), bubbles (paths that diverge then converge), frayed rope pattern (paths that converge then diverge) and cycles (paths that converge on themselves)~\cite{miller2010assembly}.
Common causes of these complexities are attributed to repeats in the target and sequencing error in the reads.
Most optimal graph reductions belong to the NP-hard class of problems, thus assemblers (like Euler, Velvet, ABySS, AllPaths, SOAPdenovo) rely on heuristics to remove redundancy, repair errors or otherwise simplify the graph.
The choice of algorithms is based on the quality, quantity and structure of the genome data.
Current short-read sequencing technologies produce very large numbers of reads favoring DBG methods.
Single molecule sequencing for third generation sequencing produces high-quality long reads which could favor OLC methods again.

In DBG the nodes represent all possible fixed-length strings of length K (K-mer graph).
The edges represent fixed-length suffix-to-prefix perfect overlaps between sub-sequences that were consecutive in the larger sequence. 
In WGS assembly, the K-mer graph represents the input reads.
Each read induces a path and those with perfect overlaps induce a common path as an advantage, however, compared to overlap graphs, K-mer graphs are more sensitive to repeats and sequencing errors as K is much less than read size.
In an ideal construction, the Eulerian path corresponding to the original sequence though graphs built from real sequencing data are more complicated. 

\begin{figure}[ht]
    \centering
    \captionsetup{justification=centering}
    \includegraphics[clip, trim=1cm 4cm 1cm 3cm, width=0.9\textwidth]{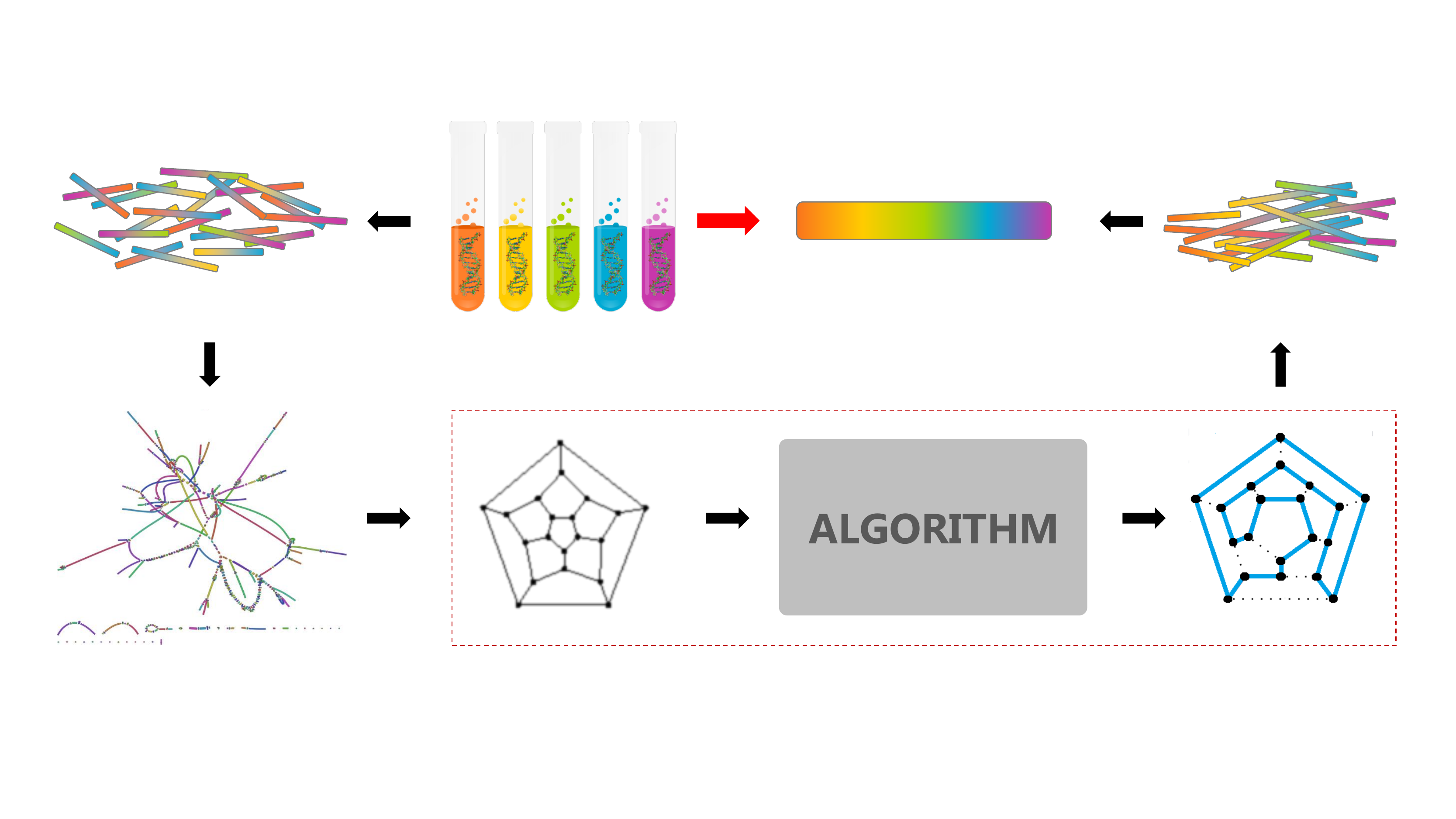}
    \caption{Overlap-Layout-Consensus genome assembly algorithm}
    \label{fig:olc}
\end{figure}

In the OLC method~\cite{commins2009computational}, as shown in figure~\ref{fig:olc}, an overlap graph represents the sequencing reads as nodes and their overlaps (pre-computed by pair-wise sequence alignments) as edges.
Paths through the graph are the potential contigs and can be converted to sequence.
Formally, it is a Hamiltonian cycle, a path that travels to every node of the graph exactly once and ends at the starting node, including each read once in the assembly.
There is no known efficient algorithm for finding a Hamiltonian cycle as it is in the NP-complete class.
Though it was feasible for microbial genome (in 1995) and the human genome (in 2001) NGS projects have abandoned it due to the computational burden.
This is the target for quantum acceleration in this research. %
\section{Quantum accelerator models} \label{s3}

There are several models of quantum computation.
The theoretical models, like the quantum circuit model, adiabatic quantum computing, measurement-based (cluster state) quantum computation and topological quantum computing are equivalent to each other within polynomial time reduction.

One of the most popular and by far the most extensively developed is the circuit model for gate-based quantum computation.
This is the conceptual generalization of Boolean logic gates (e.g. AND, OR, NOT, NAND, etc.) used for classical computation.
The gate set for the quantum counterpart allows a richer diversity of states on the complex vector space (Hilbert space) formed by qubit registers.
The quantum gates, by their unitary property, preserves the 2-norm of the amplitude of the states thereby undergoing a deterministic transformation of the probability distribution over bit strings.
The power of quantum computation stems from this exponential state space evolving in superposition while interacting by interference of the amplitudes.
Gate-based quantum algorithms are designed such that the solution state/s interfere constructively while the non-solutions interfere destructively, biasing the final probability distribution in favor of reading out the solution(s).

Another common type of a quantum system is a quantum annealer.
Quantum annealing is connected to the adiabatic quantum computing paradigm, although there are subtle differences.
While the circuit model was inspired by Boolean logic circuits, quantum annealing was inspired by the metallurgical process of annealing where by virtue of thermal fluctuations, the material is able to explore more favorable parameters (e.g. crystal size) by thermally jumping over barriers in the parameter space.
By imparting heat (energy), the solution parameters (like a ball) can climb a local minima (like a mountain and discover a deeper valley on the other side).
Quantum annealing uses quantum fluctuations instead to tunnel through high but thin barriers in the target function.
If the parameter landscape have these specific kind of barriers it can translate to a computational speedup in finding the minimum (ground state) of the function.

\begin{figure*}[ht]
    \centering
    \includegraphics[clip, trim=0cm 0cm 0cm 7cm,width=\textwidth]{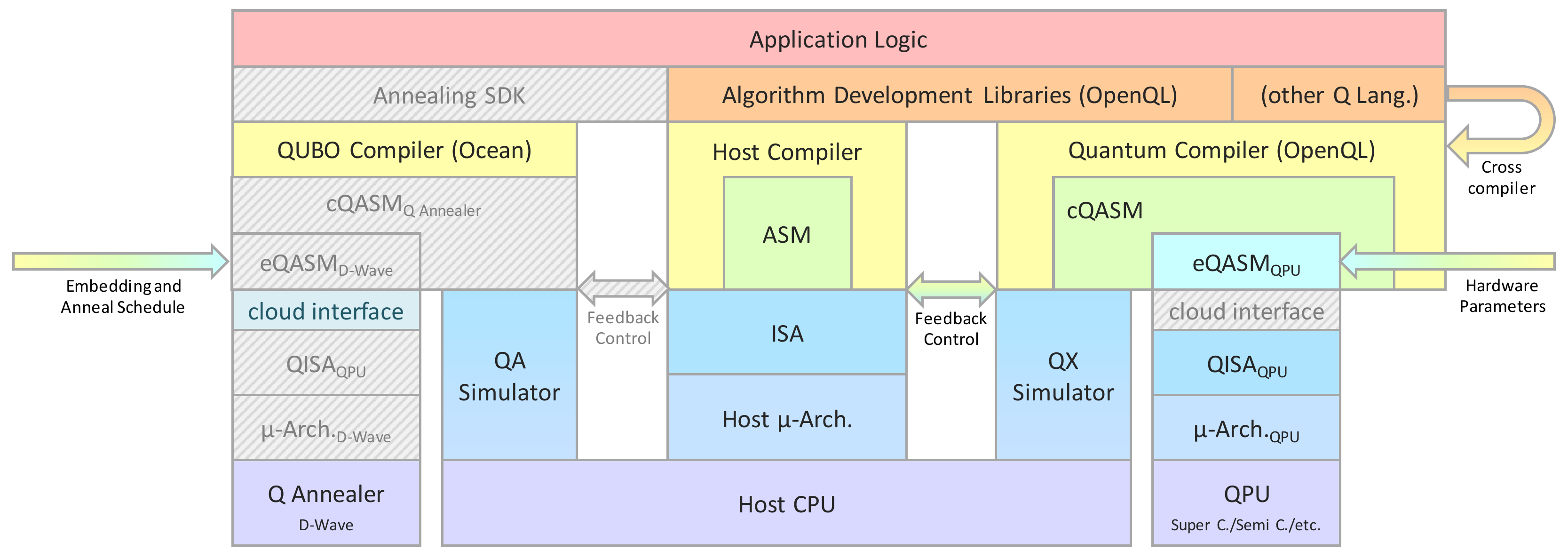}
    \caption{Quantum accelerator stack}
    \label{f_stack}
\end{figure*}

A more detailed computing stack for the execution pipeline is shown in figure~\ref{f_stack}.
The algorithms can be executed in a local simulator of the quantum accelerator before being deployed to the quantum device.
Some of the blocks 
\begin{itemize}[nolistsep,noitemsep]
    \item Application logic: The application can be described in hybrid quantum-classical logic as mathematical state evolution designed to perform the desired task. The evolution need to be decomposed into programming constructs as input to the compiler.
    \item Algorithm development libraries: To ease development of algorithms, many compilers now offer libraries which consist of an arsenal of primitives that help a quantum algorithm developer.
    \item Quantum Compiler: Quantum programming languages (like OpenQL) is the interface for the algorithm designer to precisely define the quantum operators and state in abstracted high-level constructs. For D-Wave's quantum annealer, the Ocean tools are available. The underlying blocks (in gray) are either not relevant for our discussion or proprietary. The annealer can be directly interfaced over the cloud.
    \item cQASM: In the gate-model, the compilation process generates an assembly level code (common-QASM) specifying the gate operations.
    \item Hardware parameters: Quantum runtime unit is responsible for scheduling the operations required for the compiler code. This includes quantum error correction (QEC) and qubit logical to physical mapping. The input for this is provided as hardware parameters. The executable-QASM is generated via this process. 
    \item QISA: Quantum Instruction Set Architecture (QISA) defines the runtime operations of both classical control and quantum parts of the algorithm. It encapsulates the hardware dependence.
    \item $\mu$-Arch: Quantum Micro-architecture takes into account the precise timing controls and the instruction pipelines.
    \item Simulator: Hardware agnostics application development (as discussed in this paper) can bypass to directly interface the cQASM/Ocean with the simulator (which in turn runs on the classical CPU). The simulated qubits are perfect in nature to test the functionality of the algorithm.
\end{itemize}
\section{Formulation} \label{s4}

We need to solve the minimum Hamiltonian cycle problem for the OLC method of de novo assembly.
This is also famously known as the Travelling Salesman Problem (TSP), which belongs to the NP-hard class of computational complexity.
Thus, we need to formulate it as an approximate optimization problem as even quantum computers cannot solve NP-hard problems in polynomial time.

\begin{figure}[ht]
\centering
\includegraphics[clip, trim=0cm 0cm 0cm 8cm,width=0.8\textwidth]{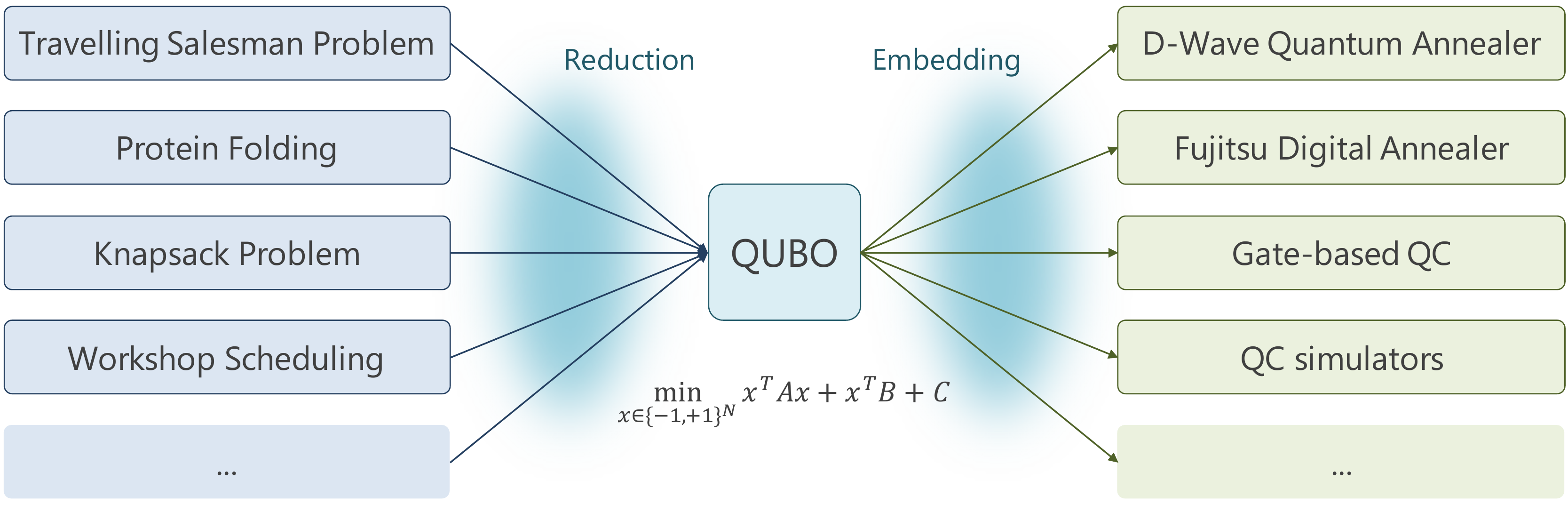}
\caption{QUBO reduction from NP-hard problems}
\label{fig:qa_qubo_redt}
\end{figure}

Mapping an NP-hard problem to quantum involves 2 steps as shown in figure~\ref{fig:qa_qubo_redt}.
The first step is to \textit{reduce} the given application to a Quadratic Unconstrained Binary Optimization (QUBO). 
The second step is to \textit{embed} the QUBO to the connectivity structure of the hardware.

\subsection{Quadratic unconstrained binary optimization}

A binary quadratic model (BQM) comprises a collection of binary-valued variables that can be assigned two chosen values (based on the model) with associated linear and quadratic biases.
Two isomorphic BQM are:
\begin{itemize}[nolistsep,noitemsep]
    \item QUBO models: $x_i \in \{0,1\}$ Boolean values
$$H(x) = \sum_iQ_{ii}x_i+\sum_{i,j}Q_{ij}x_ix_j+k$$
    \item Ising models: $\sigma_i \in \{-1,+1\}$ spin states
$$H(\sigma) = -\mu\sum_ih_i\sigma_i-\sum_{i,j}J_{ij}\sigma_i\sigma_j$$    
\end{itemize}
The choice of model depends on the problem.
Using QUBO, it might be easier to write numbers in standard binary notation (e.g. $101_2  = 5_{10}$) in the optimization problem; or it might be required to destructively interfere two variable (e.g. two battling Pokemons) using the Ising model.
Quantum processors like annealers use the Ising model, thus QUBO equations need to be converted into Ising under the hood.

QUBO model unifies a rich variety of combinatorial optimization problems as an alternative to traditional modeling and solution methodologies.
These problems are concerned with making wise choices in settings where a large number of yes/no decisions must be made and each set of decisions yields a corresponding objective function value – like a cost or profit value.
The QUBO model is expressed by the optimization problem: 
$$ \texttt{minimize } y = x^tQx = H(x)$$
where $x$ is a vector of binary decision variables and $Q$ is a (symmetric or in upper triangular) square matrix of constants.

Different types of constraining relationships arising in practice can be embodied within the unconstrained QUBO formulation using penalty functions. 
The penalties introduced are chosen so that the influence of the original constraints on the solution process can alternatively be achieved by the natural functioning of the optimizer as it looks for solutions that avoid incurring the penalties.
If the penalty terms can be driven to zero, the augmented objective function becomes the original function to be minimized.
Other than the restrictions the $x_i \in \{0,1\}$ on the decision variables, QUBO is an unconstrained model with all problem data being contained in the Q matrix.
The constrained problem: min $y = f(x)$ when subject to constraint $g(x)$ is transformed as the unconstrained form as: min $y = f(x) + P*h(g(x))$.
Penalties are not unique, meaning that many different values can be successfully employed. 
For a particular problem, a workable value is typically set, based on domain knowledge and on what needs to be accomplished.
If a constraint must be satisfied, i.e., a "hard" constraint, then $P$ must be large enough to preclude a violation. 
More moderate penalties are set for "soft" constraints, meaning that it is desirable to satisfy them but slight violations can be tolerated. 

Casting the QUBO model as a minimization problem permits a maximization problem to be solved by minimizing the negative of its objective function.
The transformation to QUBO can sometimes be aided considerably by first employing a change of variables.

QUBO models belong to the NP-hard class of problems.
Thus exact solvers (e.g. CPLEX, Gurobi) work practically only for very small problem instances (around 100 variables) using mostly branch-and-bound or problem-specific techniques.
However, impressive successes are being achieved by using meta-heuristic methods that are designed to find high quality but not necessarily optimal solutions in a modest amount of computer time.
Among the best meta-heuristic methods for QUBO are those based on tabu search, path relinking, simulated annealing, genetic/memetic strategies, and their ensembles.

\subsection{Travelling salesman problem}

A Hamiltonian path is a graph path between two vertices of a graph that visits each vertex exactly once.
If a Hamiltonian path exists whose endpoints are adjacent, then the resulting graph cycle is called a Hamiltonian cycle.
It is a path that starts from one node and ends at the same node covering all the nodes of that graph.

Given a directed complete graph $G = (V, E)$ with weights $w_{ij}$ on the directed edge $i \rightarrow j$, the directed travelling salesman problem (TSP) aims to find a directed Hamiltonian cycle of minimum weight, i.e., a cycle that visits all nodes (cities) of the graph and such that the sum of the edge weights (travel cost) is minimum.
Intuitively, given the ordered pair-wise distance between cities, the TSP involves finding the shortest route that visits every city once.
The order of visiting the cities are not constrained.

TSP falls under the NP-hard class (thus outside BQP), so the time to find the exact solution scales exponentially also on a quantum computer with the problem size.
Often a good sub-optimal solution is admissible, thus heuristic algorithms of much lesser complexity can be employed.
TSP solvers are used in many industrial applications in the domains of planning, scheduling, logistics, packing, DNA sequencing, network protocols, telescope control, VLSI testing, and many more.

The first step to specifying a TSP is to create a (weighted) graph specifying the edges in the format (vertex-from; vertex-to; weight).
Next, the TSP graph is transformed into a QUBO graph.
QUBO variables are labeled $(n, t)$ where $n$ is a node (read) in the TSP graph and $t$ is the time index of visiting it in order. E.g., if $(a, 0) = 1$ in the solution state, that means the node $a$ is visited first.
Since the total number of visits (time IDs) equals the total number of nodes (read IDs); the total possible combinations of $(n, t)$ is $|G|^2$.
$|G|$ is the number of nodes in the original TSP graph.

The QUBO graph will have $2*|G|^2*(|G|-1)$ interactions (or edges).
The interactions denote pairs of 2 nodes that can/cannot coexist.
The weight of the interaction shows the reward/penalty of coexisting.
A higher positive value denotes more penalty.
There are 3 types of penalty, for multi-location (being at 2 places at the same time), repetition (being at a city twice) and path cost for the tour.

\subsection{Hamiltonian formulation}

In physical systems (classical or quantum), a Hamiltonian describes the energy of an object.
More specifically, it describes the time-evolution of a system expressed by the Schrödinger equation:
$$ i\hbar {\frac {d}{dt}}|\psi(t)\rangle = H|\psi(t)\rangle $$
The unitary operator underlying the Hamiltonian is obtained by solving the equation for some time duration:
$U = \exp(-i Ht/\hbar)$.
The time-independent formulation of the equation reflects the total energy of the system $E = H|\psi \rangle$.

The adiabatic theorem dictates that if the change in the time-dependent Hamiltonian occurs slowly, the resulting dynamics remain simple, i.e. starting close to an eigenstate, the system remains close to an eigenstate.
For a quantum mechanical system, some initial Hamiltonian $H_i$ is slowly changed to some other final Hamiltonian $H_f$.
This implies that, if the system is started in the ground state (lowest eigenstate) of the initial Hamiltonian, the system will evolve to the ground state of the final configuration. 
The computational advantage comes from the choice of an easy-to-prepare quantum system like:
$$H_i = -\sum_i \sigma^X_i$$
(the ground state is the equal superposition state) and evolve the Hamiltonian to a system such that the ground state encodes the solution of the optimization problem we are interested in.

The change needs to be carried out by a defined schedule, for example, linear in the time scale $t\in[0,1]$ defined as: 
$$ H(t) = (1-t) H_i + t H_f $$
The energy difference between the ground state and the first excited state is called the gap, $\Delta(t)$.
If $H(t)$ has a finite gap for each $t$ during the transition the system can be evolved adiabatically with the evolution speed proportional to $1/\min(\Delta(t))^2$.
The gap, however, is highly problem-dependent, tending to have an exponentially small gap for hard problems (like those in the NP-hard class), making the time exponentially long.
Thus it is unlikely that an exact solution for these problems can be found in polynomial time.

In adiabatic quantum computation, universal calculations are performed by mapping the problem to a final Hamiltonian defined as: 
$$H_f=-\sum_{<i,j>} J_{ij} \sigma^Z_i \sigma^Z_{j} - \sum_i h_i \sigma^Z_i - \sum_{<i,j>} g_{ij} \sigma^X_i\sigma^X_j$$

Thus the system Hamiltonian $H(t)$ becomes:
$$ H(t) = (1-t) \Big[ -\sum_i \sigma^X_i \Big] + t \Big[ -\sum_{<i,j>} J_{ij} \sigma^Z_i \sigma^Z_{j} - \sum_i h_i \sigma^Z_i - \sum_{<i,j>} g_{ij} \sigma^X_i\sigma^X_j \Big] $$


The values of biases and couplings are set by the user/programmer for a quantum annealer.

The major drawback to implementing an adiabatic quantum computing directly is calculating the speed limit, which is harder than solving the original problem of finding the ground state of a Hamiltonian.
Quantum annealing drops the strict requirements of respecting speed limits in favor of repeating the transition multiple times.
Sampling from the solutions is likely to find the lowest energy state of the final Hamiltonian (though there is no theoretical guarantee).
Going from 'nearly correct' to 'correct' is still NP in general if the original problem is in NP class of complexity (the parameters for local optima aren't necessarily going to be distributed anywhere near the global optima).
However, annealing can be useful if a sub-optimal solution is acceptable for the application.

\subsubsection{QUBO to Hamiltonian: a simple example}

\begin{figure}[ht]
\centering
\includegraphics[clip, trim=17cm 0cm 0cm 9cm,width=0.5\textwidth]{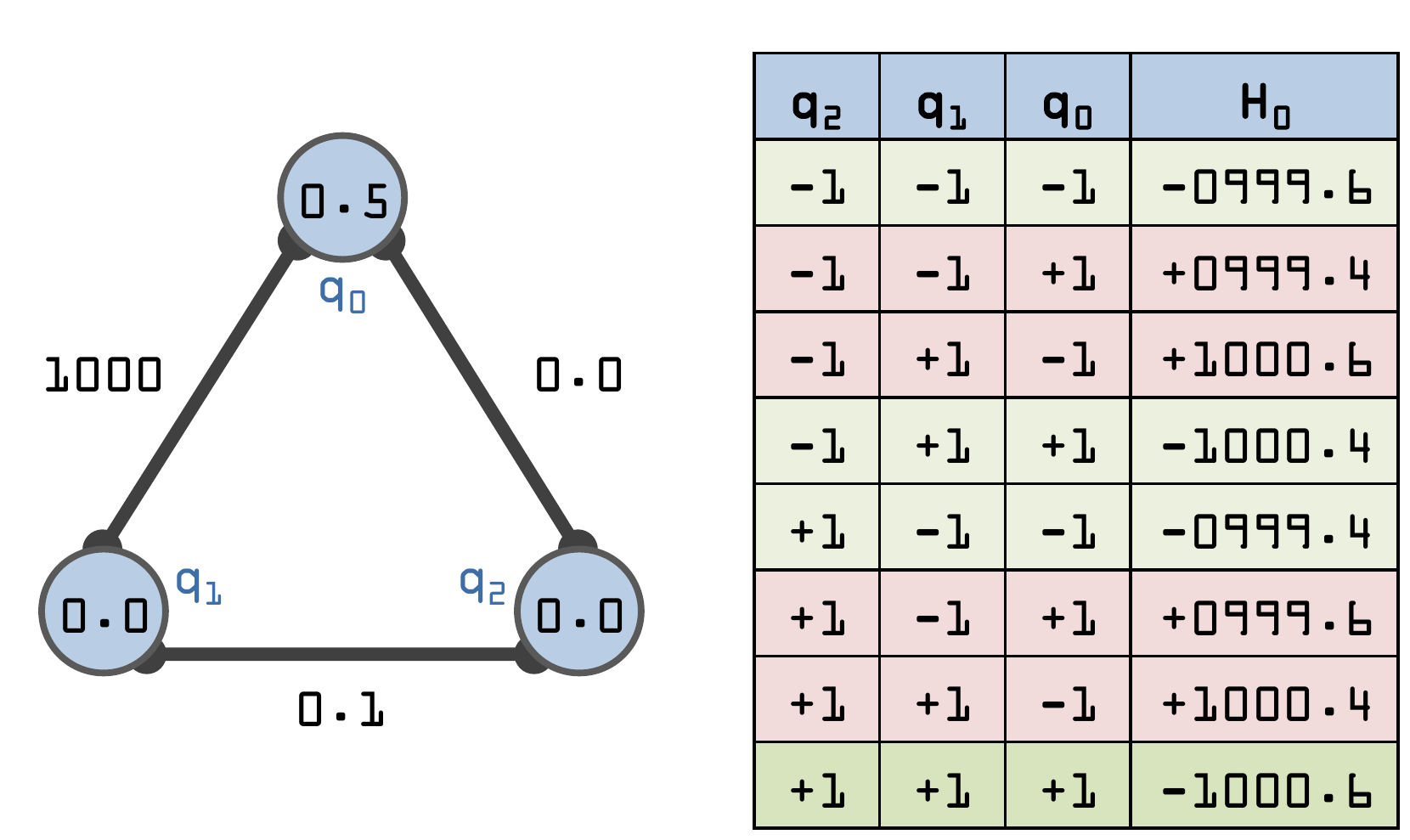}
\caption{QUBO example and possible assignments}
\label{fig:qa_qubo}
\end{figure}

Here, as an example of how to solve a QUBO on a small system of 3 qubits is shown.
In figure~\ref{fig:qa_qubo}, the biases are $[0.5,0,0]$ for qubits $\{0,1,2\}$ and the couplings for the pairs $\{(0,1),(0,2),(1,2)\}$ are $[1000,0,0.1]$ respectively.

The initial Hamiltonian is constructed with just the Pauli-X on all qubits (equal superposition):
$$H_i = -\sigma^X_0 -\sigma^X_1 -\sigma^X_2$$
The Hamiltonian encoding this graph can be constructed as:
$$H_f=-1000\sigma^Z_0\sigma^Z_1-0.1\sigma^Z_1\sigma^Z_2-0.5\sigma^Z_0$$
where $\sigma^Z$ is the Pauli-Z matrix; and all terms can be made to a canonical 3 qubit (i.e. $8\times 8$ matrix) by applying Identity to the idle qubits.

The truth table of possible assignments for the 3 qubits with values $\{-1,+1\}$ is shown in the table in figure~\ref{fig:qa_qubo}.
The lower values are more expected to be read out from the system after the optimization.
But since it is a heuristic approximate, the sub-optimal values (yellow highlights) might also be read out at times.

For this small system, it is possible to check that these two Hamiltonians have a finite gap between their two smallest eigenvalues.
Since the gap for $H_i$ is larger, the initial part of the annealing could go faster and then slow down towards reaching the target Hamiltonian.
The optimal annealing schedule is a research topic on its own.

One of the most significant applications of QUBO emerges from its equivalence to the famous Ising problem in physics.
Ising problems replace $x \in \{0,1\}^n$ by $x \in \{-1,1\}^n$ and can be put in the QUBO form by defining $x_{j'} = (x_j + 1)/2$ and then redefining $x_j$ to be $x_{j'}$, adding a constant ($1$), which is irrelevant for optimization.
The Ising problems can be solved by applying annealing methods to sample a low energy state.

\subsubsection{Using D-Wave tools to solve the Hamiltonian}

A *.qubo file lists all the constraints (all the arrows, or entries in the matrix) in a specific file format of D-Wave.
For example, the above Hamiltonian is encoded here in the file H\_example.qubo.
\begin{lstlisting}[numbers=none]    
p qubo 0 3 1 2
q0 q0 -0.5
q0 q1 -1000.0
q1 q2 -0.1
\end{lstlisting}
A parser is written to read this file into the Ising bias and couplings (or the Q matrix for the QUBO model).
\begin{lstlisting}[language=Python]  
f = open("H_example.qubo", "r")
qubo_header = f.readline().split()
hii = {}
Jij = {}
Q = {}
for i in range(0,int(qubo_header[4])):
	x = f.readline().split()
	hii[x[0]] = float(x[2])	
	Q[(x[0],x[1])] = float(x[2])	
for i in range(0,int(qubo_header[5])):
	x = f.readline().split()
	Jij[(x[0],x[1])] = float(x[2])
	Q[(x[0],x[1])] = float(x[2])
f.close()
\end{lstlisting}
For solving the QUBO/Ising model, D-Wave System's dimod package is used that simulates the behavior of their Quantum Annealer.
It provides a binary quadratic model (BQM) class that contains Ising and QUBO models used by samplers such as the D-Wave system. 
Samplers are processes that sample from low-energy states of a problem's objective function. 
D-Wave provides samplers~\cite{ocean} for running on classical systems like exact solver, null sampler, random sampler, simulated annealing sampler.
The exact solver of D-Wave's dimod package can give exhaustive evaluations for $\le 20$ variables.
\begin{lstlisting}[language=Python]  
import dimod
solver = dimod.ExactSolver()
response = solver.sample_ising(hii, Jij)
for sample, energy in response.data(['sample', 'energy']): print(sample, energy)
\end{lstlisting}
The output matches with the table in figure~\ref{fig:qa_qubo}. Similarly, the QUBO version can be solved using solver.sample\_qubo(Q) function with the Q-matrix instead.
\begin{lstlisting}[language=bash,numbers=none]  
{'q0':  1, 'q1':  1, 'q2':  1} -1000.6
{'q0':  1, 'q1':  1, 'q2': -1} -1000.4
{'q0': -1, 'q1': -1, 'q2': -1}  -999.6
{'q0': -1, 'q1': -1, 'q2':  1}  -999.4
{'q0':  1, 'q1': -1, 'q2': -1}   999.4
{'q0':  1, 'q1': -1, 'q2':  1}   999.6
{'q0': -1, 'q1':  1, 'q2':  1}  1000.4
{'q0': -1, 'q1':  1, 'q2': -1}  1000.6
\end{lstlisting}

\subsection{Variational hybrid approach}

Coherent quantum protocols have promising exponential speedups but assumes a fault-tolerant quantum computing (FTQC) platform, with high quality of qubits, a large number of gates and circuit width.
These popular algorithms like Shor's factorization, HHL for matrix inversion, though primitives for many quantum algorithms are not of immediate practical relevance.
Wrapping these protocols in state preparation (from classical data to quantum) and state tomography (from quantum probability amplitudes of final state to classical statistics) can overrule the entire speedup achieved by the protocol itself.

Today, we are in the Noisy Intermediate-Scale Quantum (NISQ) era as coined by John Preskill.
These near-term quantum computers are more suited for hybrid quantum-classical (HQC) algorithms~\cite{cao2018quantum}.
These are heuristic protocols based on the variational principle.
In an HQC algorithm, all the power of the quantum computer is used for preparing a quantum state.
The complexity of the algorithm is traded-off for multiple measurements over multiple cycles.
The operations that require lots of gates on a quantum computer are offloaded to the classical computer (e.g. optimization, addition, division), which controls the quantum computer like an accelerator or a co-processor.
A quantum circuit is defined as having a certain format $A$ (or ansatz/stencil) with parameters.
There are $m$ parameters forming a parameter vector $\Lambda_m$.
These can be initialized randomly or with a classical guess.
For the first cycle, the quantum computer takes the initial guess and evolves it using the circuit $A(\Lambda_m^0)$.
The Hamiltonian (energy) is measured out and sent to the classical computer.
The variational principle updates the parameters in such a way that the energy of the Hamiltonian is lowered in each successive iteration.
The optimization using $\Lambda_m^1,\Lambda_m^2,\Lambda_m^3,\dots$ continues until the acceptable threshold is satisfied, very similar to training in neural networks.

\begin{figure}[ht]
\centering
\includegraphics[clip, trim=12cm 0cm 0cm 11cm,width=0.9\textwidth]{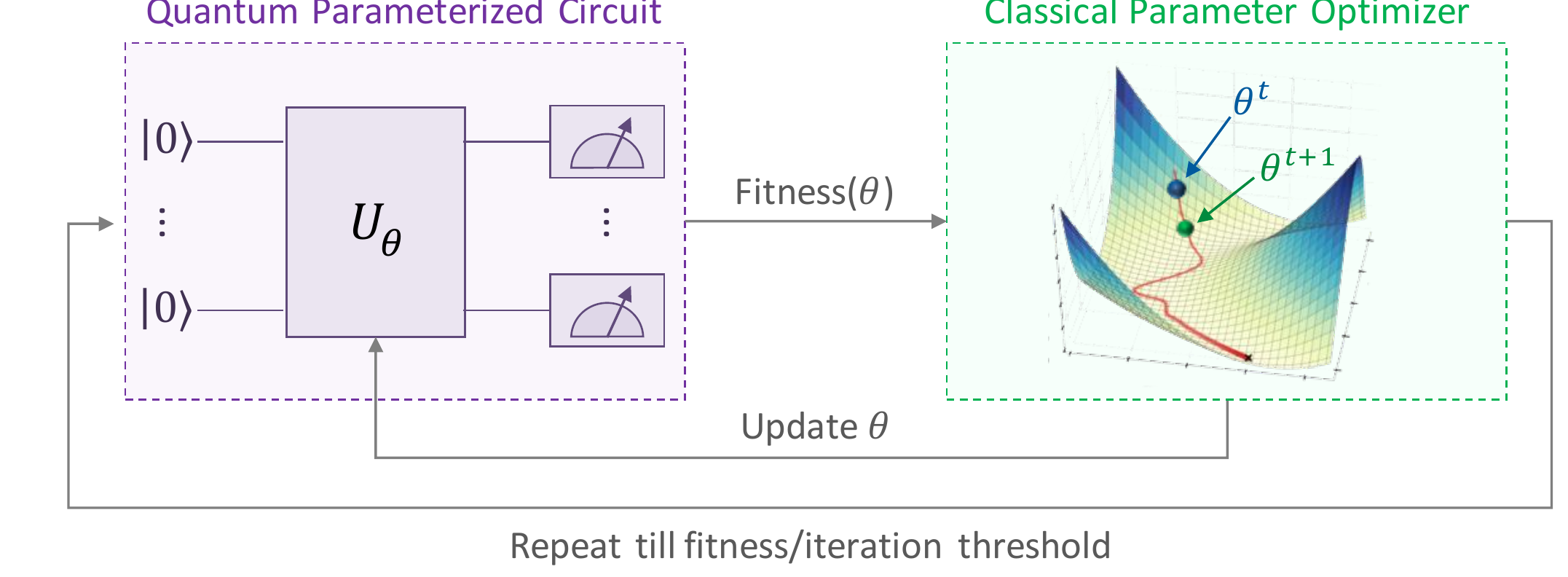}
\caption{Variational hybrid quantum-classical approach for optimization}
\label{fig:hqc}
\end{figure}

The variational principle forms the core theoretical basis behind the working of near-term quantum heuristic algorithms.
It states that, for a trial wave-function (defining the family of quantum states reachable by varying the $m$ parameters of $A$),
$$ E_{\Psi_T} = \dfrac{\bra{\Psi_T(\Lambda)} \hat{H}_{QC} \ket{{\Psi_T(\Lambda)}}}{\braket{\Psi_T(\Lambda)|\Psi_T(\Lambda)}} \geq E_0 $$
The normalization term $\braket{\Psi_T(\Lambda)|\Psi_T(\Lambda)} = 1$ as we assume no leakage errors from the computational basis. 
Thus, it is possible to reach the ground-state energy by finding the right parameters.
The more free the state is to represent quantum states (determined by the choice of $A$), the better it will be able to lower the energy.

Since adiabatic and gate systems offer effectively the same potential for achieving the gains inherent in quantum computing processes, analogous advances associated with QUBO models may ultimately be realized through quantum circuit systems as well.

An example of HQC algorithms is the Quantum Approximate Optimization Algorithm (QAOA)~\cite{farhi2014quantum}.
It is a hybrid variational algorithm that produces approximate solutions for combinatorial optimization problems.
In theory, QAOA methods can be applied to more types of combinatorial optimization problems than embraced by the QUBO model~\cite{glover2018tutorial}.
The parameters of the QAOA framework must be modified to produce different algorithms to appropriately handle different problem types.
QAOA is a polynomial-time HQC algorithm 
which can be seen as the Trotterization of infinite time adiabatic algorithm. 
Since the AQC always gives the optimal solution for Hamiltonians with a non-zero gap, QAOA for infinite cycles also converges to the global optima. 


The generalization of QAOA called the Quantum Alternating Operator Ansatz~\cite{hadfield2019quantum}, consists of 2 Hamiltonians: a cost/problem Hamiltonian $H_C$ (similar to the transverse field in AQC) and a driver/mixing Hamiltonian $H_M$ (similar to the longitudinal field in AQC).
This is repeated over $p$ cycles with each Hamiltonian parameterized by the $\gamma$ and $\beta$ real values (rotation angles similar to the adiabatic evolution time).
After this unitary evolution, the state is measured for the expectation value with respect to the ground state of the cost Hamiltonian.
The initial state $\ket{\psi_0}$ depends on the problem (typically either the all-zero or the equal superposition state).
$$U(\theta)\ket{\psi_0} = H_M(\beta_p)H_C(\gamma_p)\dots H_M(\beta_2)H_C(\gamma_2)H_M(\beta_1)H_C(\gamma_1)\ket{\psi_0}$$

For an optimization instance, the user specifies the driver Hamiltonian ansatz, cost Hamiltonian ansatz, and the approximation order (cycles) of the algorithm.
If the number of cycles in QAOA increases, theoretically, the sub-optimal solutions obtained can only get better, as the sub-optimal solutions defined by fewer cycles (with fewer free parameters) are always contained in more cycles (if the new rotation parameters are set to zero).
However, practically, having more cycles causes difficulty for the classical optimizer to deal with more free parameters and can affect the convergence.

Since HQC trades off the decoherence issue of a long quantum circuit in the NISQ era with multiple low-depth, the number of repetitions required is high.
To estimate the expectation of the prepared state in each optimization step, it needs to be (pre-rotated in the basis and) measured with respect to each Pauli term in the problem Hamiltonian and aggregated.
Each Pauli term measurement in turn requires state tomographic trials.

Moreover, the optimizer might get stuck in local optima or barren plateaus~\cite{mcclean2018barren} in the parameter landscape, requiring a few reruns to build confidence in the obtained optima.
The HQC algorithms depend a lot on the choice of the classical optimizer as well.
Here we experiment with the basic Nelder-Mead gradient-free optimizer, but many gradient-based and gradient-free choices exist (for example in libraries like SciPy in Python and TOMLAB in MATLAB) which needs to be chosen based on empirical testing of a particular formulation of the specific problem.

The pseudo-code for the OpenQL implementation of the QAOA algorithm is shown in Listing~\ref{code:pqc}.
\begin{lstlisting}[numbers=none,columns=fullflexible,backgroundcolor=\color{codeblue},mathescape=true,frame=lines,caption=Pseudo-code for OpenQL QAOA,label={code:pqc}]
$\bullet$ Invoke QAOA:
    $\diamond$ Form parameterized cQasm [Input: reference circuit, ansatz, steps, coefficients, angle ids]
    $\diamond$ Form parameters list [Input: gammas, betas]
    $\diamond$ Set runtime deparameterized cQasm filename
    $\diamond$ For each iteration:
        $\star$ For each Pauli product term in cost Hamiltonian (wsopp):
            $\circ$ Deparameterize cQasm [Input: parameterized QASM, parameter list]
            $\circ$ Add measurement basis rotation based on Pauli product term
            $\circ$ Invoke Qxelerator to aggregate measurement over shots
        $\star$ Calculate total Expectation value of trail state in cost Hamiltonian basis
        $\star$ Return value to optimizer. Save intermediate results via callback.
        $\star$ Classical optimizer updates parameter list
    $\diamond$ Display cost function convergence, final parameters and optimized cost
\end{lstlisting} %
\section{Implementing de novo assembly} \label{s5}

In this section, a fully worked out example is presented as a proof-of-concept for quantum accelerated sequence reconstruction.
First, we will present a mapping from DNA to TSP.
Then the TSP will be converted to a QUBO.

\subsection{DNA reads to TSP formulation}\label{s5p1}

Suppose the sequencer produces reads of size $10$ and after removing duplicates, the reads obtained are:\\
\textit{read 0 : ATGGCGTGCA}\\
\textit{read 1 : GCGTGCAATG}\\
\textit{read 2 : TGCAATGGCG}\\
\textit{read 3 : AATGGCGTGC}

The pairwise overlap is calculated for each ordered pair based on how many prefix characters of the second string match exactly with the suffix of the first string.
The edge weight is set to the negation of the overlap, as the constraints need to be formulated such that the path that \textit{minimizes the overlap} is found.

The edge weights are \emph{\{(0,1):-7 , (1,2):-7 , (2,3):-7 , (3,0):-9, (1,0):-3 , (2,1):-3 , (3,2):-3 , (0,3):-1, (0,2):-4 , (1,3):-4 , (2,0):-6 , (3,1):-6\}}

\begin{figure}[ht]
\centering
\includegraphics[clip, trim=0cm 0cm 0cm 0cm,width=\textwidth]{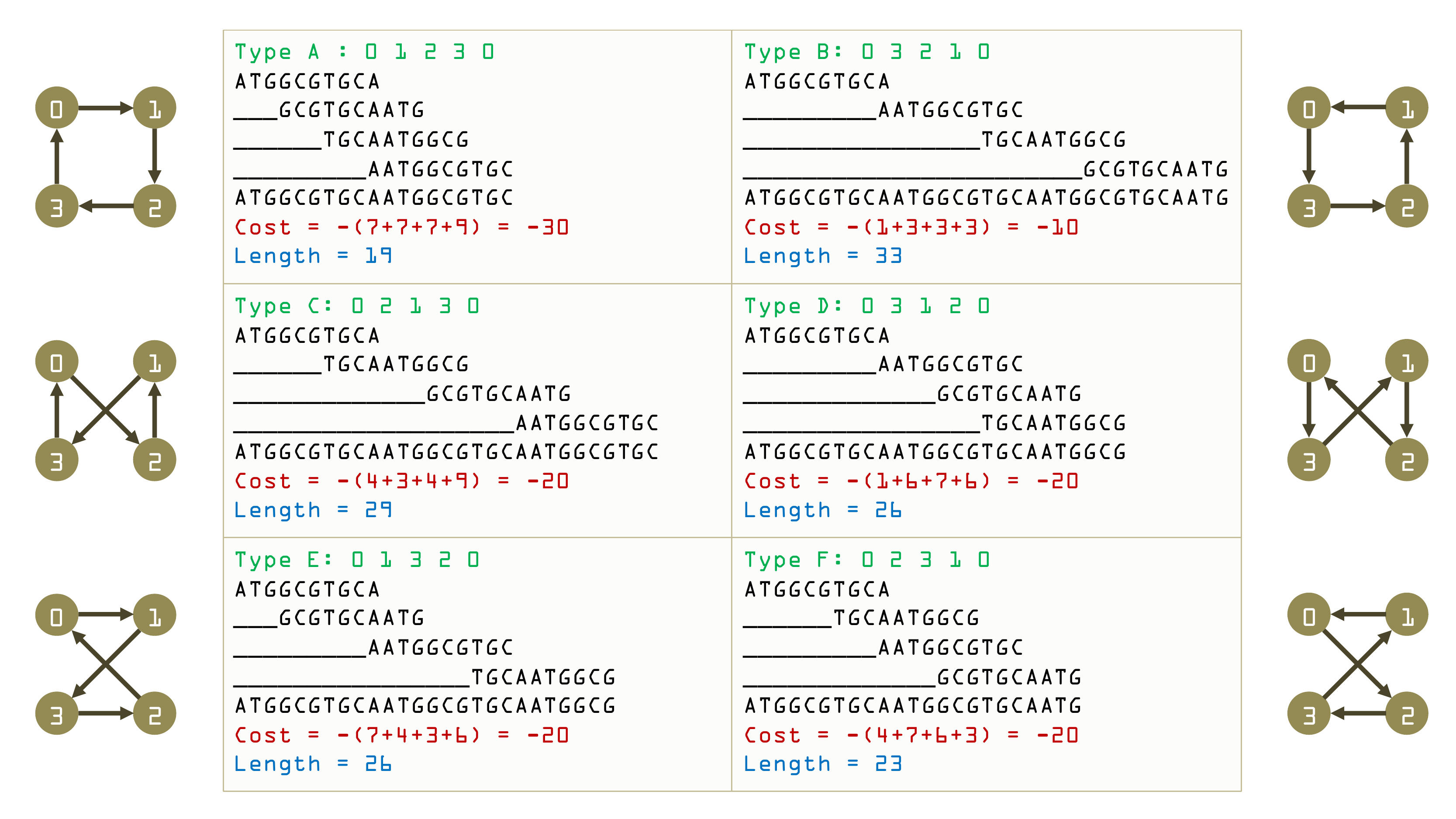}
\caption{All possible TSP tours for given example}
\label{fig:paths}
\end{figure}

The overlap depends on the ordering of the read pairs and thus this formulation is a directed graph.
There are 6 possible unique tours (choosing a different starting city in the tour is equivalent in cost).
Note that the reads are spliced from an original circular DNA, so the final stitched DNA solutions for all these tours are repeats of \emph{read 0} of variable length.
Such cases occur in practice when arranging the reads in different ways gives different repeat lengths. 
The tours are emulated in figure~\ref{fig:paths}.
The TSP solution is expected to find the lowest cost (and shortest assembly) tour, i.e. Type-A.
There are 4 acceptable solutions (based on starting node) of Type-A:
\begin{itemize}[nolistsep,noitemsep]
    \item $(0 \rightarrow 1 \rightarrow 2 \rightarrow 3 \rightarrow 0)$
    \item $(1 \rightarrow 2 \rightarrow 3 \rightarrow 0 \rightarrow 1)$
    \item $(2 \rightarrow 3 \rightarrow 0 \rightarrow 1 \rightarrow 2)$
    \item $(3 \rightarrow 0 \rightarrow 1 \rightarrow 2 \rightarrow 3)$
\end{itemize}

This process is automated in the following classical pre-processing script.
\begin{lstlisting}[language=Python] 
# Overlap between pair-wise reads
def align(read1,read2,mm):
	l1 = len(read1)
	l2 = len(read2)
	for shift in range(l1-l2,l1):
		mmr = 0
		r2i = 0
		for r1i in range(shift,l1):
			if read1[r1i]!=read2[r2i]:
				mmr += 1
			r2i += 1
			if mmr > mm:
				break
		if mmr <= mm:
			return l2-shift
	return 0

# Convert set of reads to adjacency matrix of pair-wise overlap for TSP
def reads_to_tspAdjM(reads, max_mismatch = 0):
	n_reads = len(reads)
	O_matrix = np.zeros((n_reads,n_reads)) # edge directivity = (row id, col id)
	for r1 in range(0,n_reads):
		for r2 in range(0,n_reads):
			if r1!=r2:
				O_matrix[r1][r2] = align(reads[r1],reads[r2],max_mismatch)
	O_matrix = O_matrix / np.linalg.norm(O_matrix)
	return O_matrix

reads = ['ATGGCGTGCA','GCGTGCAATG','TGCAATGGCG','AATGGCGTGC']
tspAdjM = reads_to_tspAdjM(reads)
\end{lstlisting}

\subsection{TSP to QUBO model}

Next, the directed TSP is encoded as a QUBO model.
Let $n = |V| = 4$ be the number of nodes.
This formulation~\cite{nannicini2019performance} requires $n^2 = 16$ binary variables as qubits, so it scales quadratically rather than linearly in the problem size.
For $i, p \in \{0 \dots (n-1)\}$, let $x_{i,p}$ be True if node $i$ appears in position $p$ in the cycle, False otherwise.

To derive a Hamiltonian for this problem, we penalize the violation of the constraints in the objective function inserting terms of the form $\alpha (\sum_{p=0}^{n-1} x_{i,p} - 1)^2$, where the penalty term is sufficiently large, e.g., $\alpha = n * max_{(i,j)\in E} w_{ij}$.
The TSP can be formulated as: $$\forall i,p \in \{0 \dots (n-1)\}, x_{i,p}\in\{0,1\}$$

\begin{figure}[ht]
\centering
\includegraphics[clip, trim=22cm 0cm 0cm 22cm,width=\textwidth]{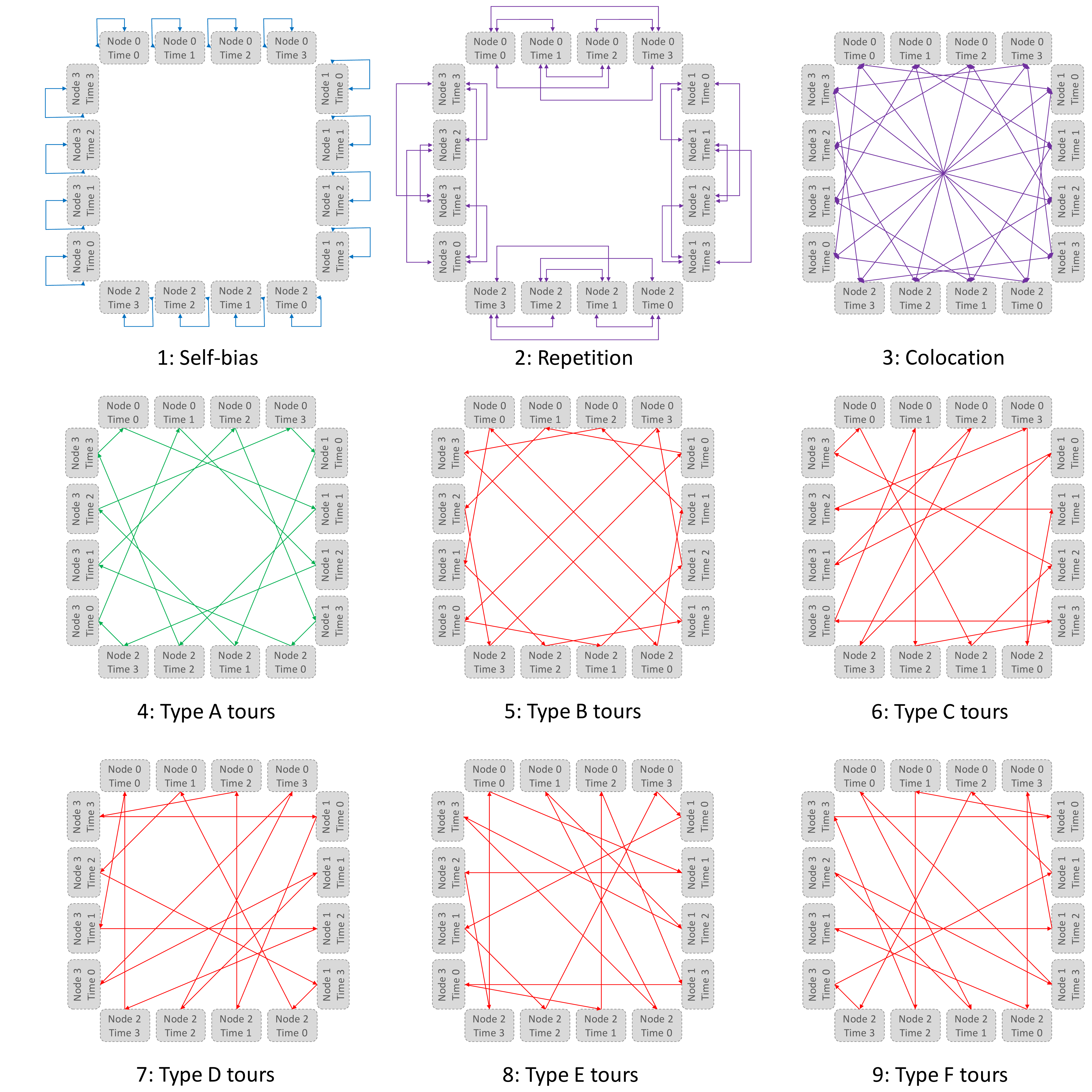}
\caption{QUBO interactions}
\label{fig:qa_tsp_dna}
\end{figure}

The interactions are shown in figure~\ref{fig:qa_tsp_dna} and are categorized as:
\begin{enumerate}[nolistsep,noitemsep]
    \item Every node must be assigned. Thus self-interactions have large negative weight (favorable bias). Since there are no preferred order of the route, for each time slot, the value is the same (top-left blue interactions).
    \item Same node assigned to two different time slots incurs a penalty (top-middle violet interactions). Thus, for each node, there should be only one assigned time slot:
    $$\forall i \in \{0 \dots (n-1)\}, \sum_{p=0}^{n-1} x_{i,p} = 1$$
    \item Same time slot assigned to two different nodes incurs a penalty (top-right violet interactions). Thus, for each time slot, there should be only one assigned node:
    $$\forall p \in \{0 \dots (n-1)\}, \sum_{i=0}^{n-1} x_{i,p} = 1$$
    \item The additional cost of including an edge in the route to two consecutive time slots is the weight of the edge in the TSP. These 6 graphs show the 4 possible routes for each of the 6 types (type A in middle-left green, others in red). 
    Each edge $(i,j)$ is taken and all possible configurations of assigning them next to each other are tried (the addition being modulo $n$), with the edge weight being the cost of choosing from those configurations.
    Thus, given the above constraints:
    $$ \texttt{ minimize: }\sum_{i=0}^{n-1} \sum_{j=0}^{n-1} w_{ij} \sum_{p=0}^{n-1} x_{i,p}x_{j,p+1}$$ 
    
\end{enumerate}

\begin{figure}[htbp]
\centering
\includegraphics[clip, trim=19cm 0cm 0cm 4cm,width=0.95\textwidth]{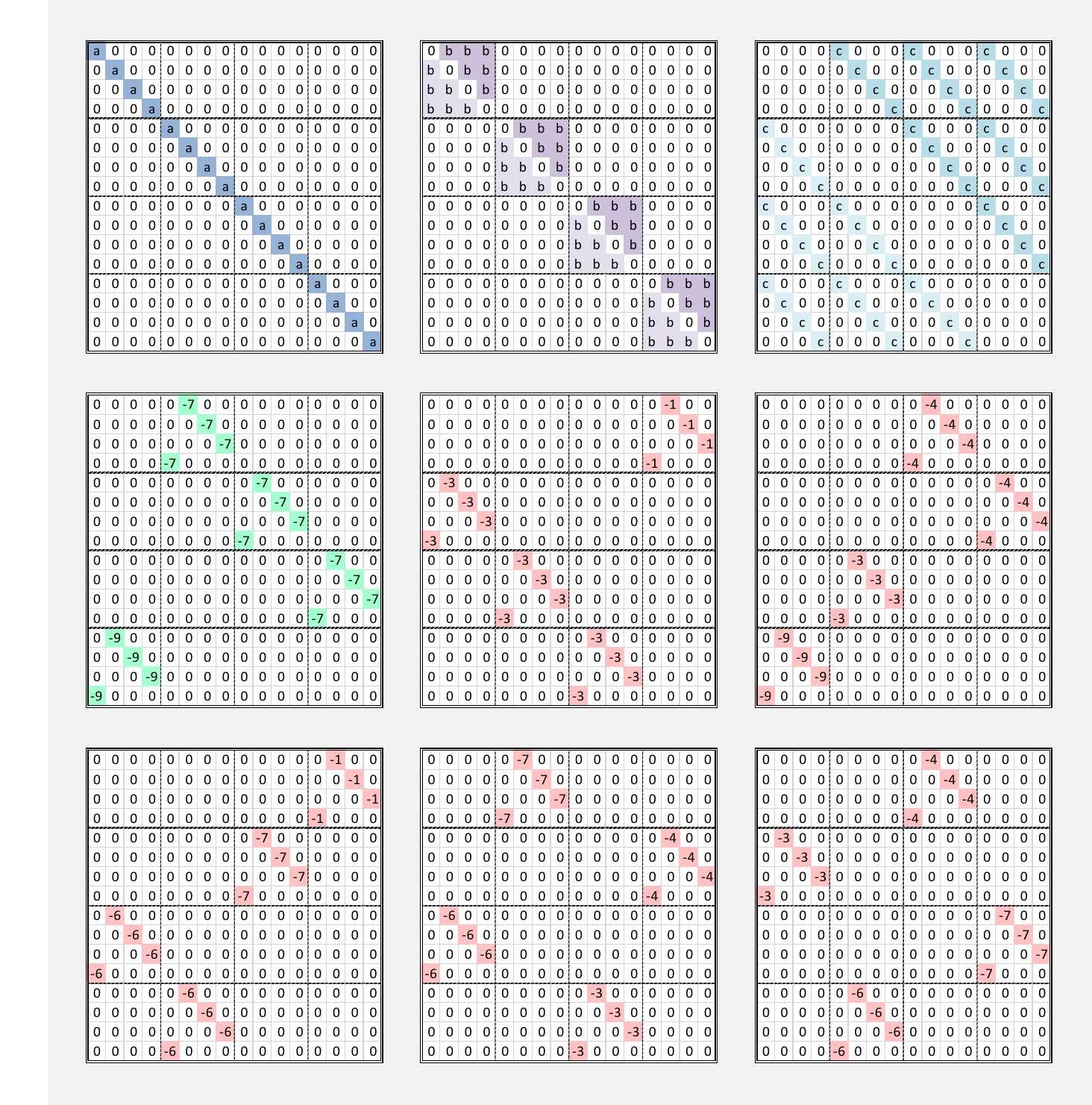}
\caption{QUBO interactions Q-matrix}
\label{fig:qa_tspmatrix_dna}
\end{figure}

These arrows can be made into an adjacency matrix for the graph, as shown in figure~\ref{fig:qa_tspmatrix_dna}.
The addition of these 6 matrices gives the $Q$ matrix for the QUBO.
Note that, if we have only the lower 6 matrices (coupling), the all 1's assignment is the most favorable and gives the minimum solution to the QUBO equation $y = x^TQx$.
Thus, we need to add the reward for assigning a node $\{a\}$ and penalties $\{b,c\}$ for assigning the same node to multiple different time slots, or same time slots to multiple nodes, respectively.
Since these are bi-directional arrows, these can be symmetric.

For our experiment, we empirically found that setting a penalty value of $b = c \ge 13 $ is sufficient, and the reward $a = 0$ still finds the 4 favorable minima of Type A.
It is easy to verify that a minimum value of $y = x^TQx$ is obtained for:
\begin{itemize}[nolistsep,noitemsep]
    \item $x^T = [1 0 0 0 0 1 0 0 0 0 1 0 0 0 0 1]$
    \item $x^T = [0 1 0 0 0 0 1 0 0 0 0 1 1 0 0 0]$
    \item $x^T = [0 0 1 0 0 0 0 1 1 0 0 0 0 1 0 0]$
    \item $x^T = [0 0 0 1 1 0 0 0 0 1 0 0 0 0 1 0]$
\end{itemize}
for the binary encoding 
$$x^T = [n_0t_0 | n_0t_1 | n_0t_2 | n_0t_3 | n_1t_0 | n_1t_1 | n_1t_2 | n_1t_3 | n_2t_0 | n_2t_1 | n_2t_2 | n_2t_3 | n_3t_0 | n_3t_1 | n_3t_2 | n_3t_3]$$

The process of generating the Q matrix is automated in the following classical pre-processing script.

\begin{lstlisting}[language=Python] 
# Convert adjacency matrix of pair-wise overlap for TSP to QUBO matrix of TSP
def tspAdjM_to_quboAdjM(tspAdjM, p0, p1, p2):
	n_reads = len(tspAdjM)
	# Initialize
	Q_matrix = np.zeros((n_reads**2,n_reads**2)) # Qubit index semantics: {c(0)t(0) |..| c(i)-t(j) | c(i)t(j+1) |..| c(i)t(n-1) | c(i+1)t(0) |..| c(n-1)t(n-1)}
	# Assignment reward (self-bias)
	p0 = -1.6
	for ct in range(0,n_reads**2):
		Q_matrix[ct][ct] += p0
	# Multi-location penalty
	p1 = -p0 # fixed empirically by trial-and-error
	for c in range(0,n_reads):
		for t1 in range(0,n_reads):
			for t2 in range(0,n_reads):
				if t1!=t2:
					Q_matrix[c*n_reads+t1][c*n_reads+t2] += p1
	# Visit repetition penalty
	p2 = p1
	for t in range(0,n_reads):
		for c1 in range(0,n_reads):
			for c2 in range(0,n_reads):
				if c1!=c2:
					Q_matrix[c1*n_reads+t][c2*n_reads+t] += p2
	# Path cost
	# kron of tspAdjM and a shifted diagonal matrix
	for ci in range(0,n_reads):
		for cj in range(0,n_reads):
			for ti in range(0,n_reads):
				tj = (ti+1)%n_reads
				Q_matrix[ci*n_reads+ti][cj*n_reads+tj] += -tspAdjM[ci][cj]
	return Q_matrix

quboAdjM = tspAdjM_to_quboAdjM(tspAdjM, 0, 13, 13) # self-bias, multi-location, repetation
\end{lstlisting} %
\section{Solving on a quantum system} \label{s6}

In this section, we show how to solve the QUBO on both a quantum annealer (D-Wave Ocean tools) and a gate-based optimizer (QX/OpenQL) using QAOA.

\subsection{QUBO using quantum annealing}

The QUBO is mapped to a quantum annealer using the biases and couplings in the Ising model.
A bias value is defined for each qubit and a coupling for each pair of qubits.
In the graph view, each node (bias) and each edge (coupling) can have a real-number value (weight).
Since this example requires 16 QUBO variables (qubits), the exact solver is used to better understand the output.

\begin{figure}[ht]
\centering
\includegraphics[clip, trim=20cm 0cm 0cm 5cm,width=0.4\textwidth]{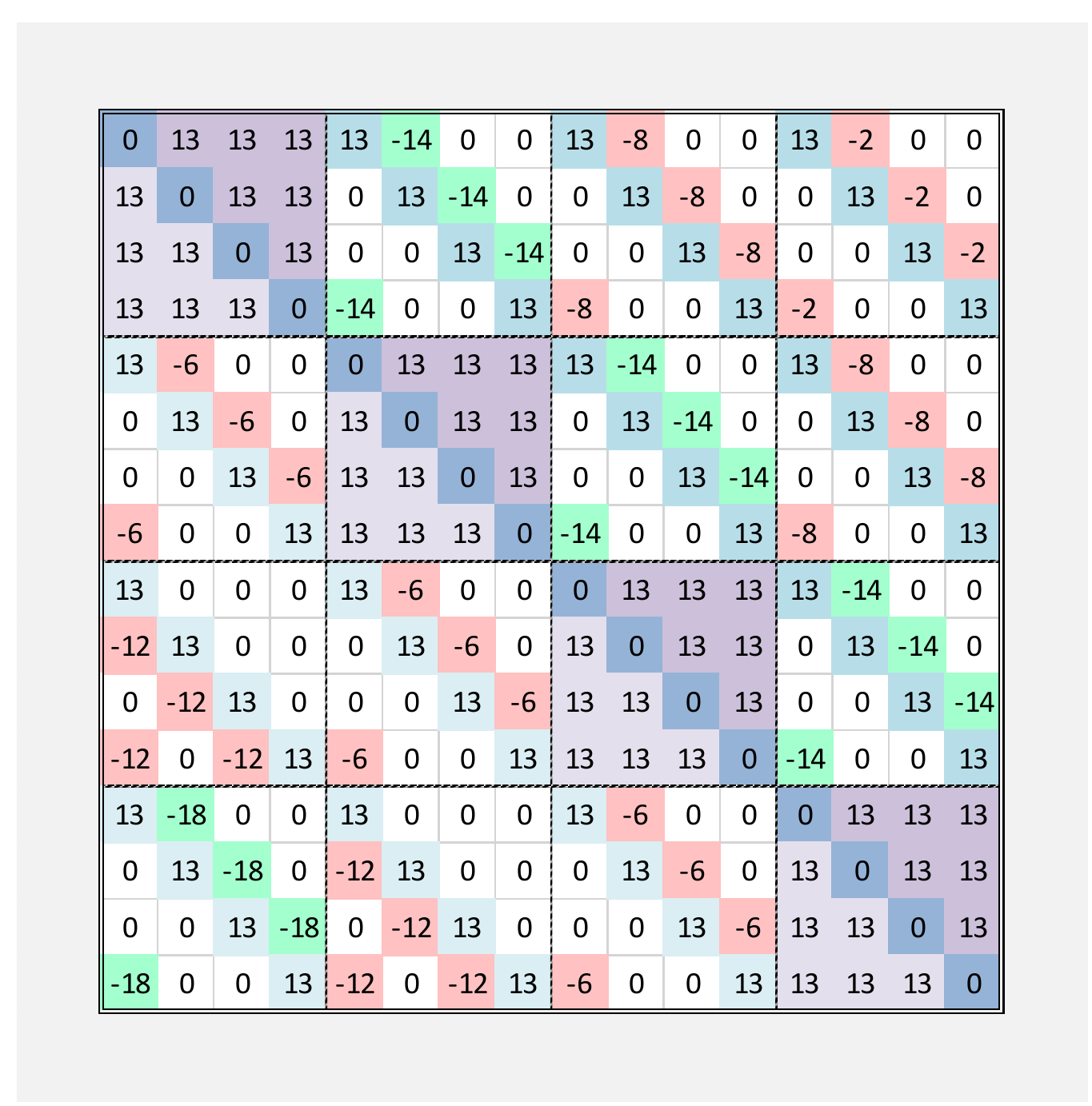}
\caption{Q-matrix for de novo example}
\label{fig:qa_qm_dna}
\end{figure}

For the de novo example, the Q matrix from figure~\ref{fig:qa_tspmatrix_dna} (with $a = 0, b=c=13$) is shown in figure~\ref{fig:qa_qm_dna}.

The Q matrix is converted to a dictionary of node names and reward/penalty for biases and couplings.
Then the QUBO solver is used to solve the Q matrix using the assignment of $\{0,1\}$ (instead of the Ising $\{-1,+1\}$).
\begin{lstlisting}[language=Python]
# Convert QUBO matrix of TSP to QUBO dictionary of weighted adjacency list
def quboAdjM_to_quboDict(Q_matrix):
	n_reads = int(math.sqrt(len(Q_matrix)))
	Q = {}
	for i in range(0,n_reads**2):
		ni = 'n'+str(int(i/n_reads))+'t'+str(int(i%n_reads))
		for j in range(0,n_reads**2):
			nj = 'n'+str(int(j/n_reads))+'t'+str(int(j%n_reads))
			if Q_matrix[i][j] != 0:
				Q[(ni,nj)] = Q_matrix[i][j]
	return Q
	
# Solve a QUBO model using dimod exact solver
import dimod
def solve_qubo_exact(Q, all = False):
	solver = dimod.ExactSolver()
	response = solver.sample_qubo(Q)
	minE = min(response.data(['sample', 'energy']), key=lambda x: x[1])
	for sample, energy in response.data(['sample', 'energy']): 
		if all or energy == minE[1]:
			print(sample)
			
Q = quboAdjM_to_quboDict(Q_matrix)
solve_qubo_exact(Q)
\end{lstlisting}
We find that there are 4 minimum solutions (the 4 Type A solutions). 
This matches with the expected analytical results.





Though the QUBO solution now works on D-Wave's solver, it needs to be converted to the Ising model for it to run on the D-Wave Quantum Annealer.
This can easily be done using the qubo\_to\_ising function in D-Wave's toolset, which maps the definitions of binary variables to an Ising model defined on spins (variables with {-1, +1} values).
The following script solves the Ising model for our formulation.

Mathematically, the transform is: $x^TQx = \text{offset} + s^TJs + h^Ts$.
For every linear (diagonal) bias term in Q, $h[i] += 0.5 * Q[i][i]$, while for each couplings, $J[(i, j)] = 0.25 * Q[i][j]$, $h[i] += 0.25 * Q[i][j]$, $h[j] += 0.25 * Q[i][i]$.
The offset value is the weighted sum of the linear offset (sum of all diagonal terms in Q), and the quadratic offset (sum of all off-diagonal terms in Q), with the weights as $0.5$ and $0.25$, respectively.
The offset is not important for our case as we want the qubit state of the minimum energy, not the exact value of the minimized energy.
\begin{lstlisting}[language=Python] 
# Solve an Ising model using dimod exact solver
import dimod
def solve_ising_exact(hii,Jij, plotIt = False):
	solver = dimod.ExactSolver()
	response = solver.sample_ising(hii,Jij)
	print("Minimum Energy Configurations\t===>")
	minE = min(response.data(['sample', 'energy']), key=lambda x: x[1])
	for sample, energy in response.data(['sample', 'energy']): 
		if energy == minE[1]:
			print(sample,energy)
			
hii, Jij, offset = dimod.qubo_to_ising(Q)		
solve_ising_exact(hii,Jij)
\end{lstlisting}
As expected, we find that there are 4 minimum solutions (the 4 Type A solutions).
\begin{lstlisting}[language=bash,numbers=none]
{'n0t0': +1, 'n0t1': -1, 'n0t2': -1, 'n0t3': -1, 
 'n1t0': -1, 'n1t1': +1, 'n1t2': -1, 'n1t3': -1, 
 'n2t0': -1, 'n2t1': -1, 'n2t2': +1, 'n2t3': -1, 
 'n3t0': -1, 'n3t1': -1, 'n3t2': -1, 'n3t3': +1}

{'n0t0': -1, 'n0t1': +1, 'n0t2': -1, 'n0t3': -1, 
 'n1t0': -1, 'n1t1': -1, 'n1t2': +1, 'n1t3': -1, 
 'n2t0': -1, 'n2t1': -1, 'n2t2': -1, 'n2t3': +1, 
 'n3t0': +1, 'n3t1': -1, 'n3t2': -1, 'n3t3': -1}

{'n0t0': -1, 'n0t1': -1, 'n0t2': +1, 'n0t3': -1,
 'n1t0': -1, 'n1t1': -1, 'n1t2': -1, 'n1t3': +1, 
 'n2t0': +1, 'n2t1': -1, 'n2t2': -1, 'n2t3': -1, 
 'n3t0': -1, 'n3t1': +1, 'n3t2': -1, 'n3t3': -1}

{'n0t0': -1, 'n0t1': -1, 'n0t2': -1, 'n0t3': +1,
 'n1t0': +1, 'n1t1': -1, 'n1t2': -1, 'n1t3': -1, 
 'n2t0': -1, 'n2t1': +1, 'n2t2': -1, 'n2t3': -1, 
 'n3t0': -1, 'n3t1': -1, 'n3t2': +1, 'n3t3': -1}
\end{lstlisting}

\subsubsection{Using D-Wave Quantum Annealer}

D-Wave offers a connection to the cloud to solve an Ising model.
The subscription token and preferred system is part of a configuration file.
Once the solver of a quantum annealer on the cloud is selected, the connectivity of the system can be retrieved.

Then the problem needs to be embedded in the graph.
It is called a Chimera graph for the D-Wave 2000Q systems.
Each of the 16 logical qubits is embedded over multiple qubits on the actual hardware so that each qubit shares a coupling based on the required interaction for the Ising model.
The embedding process is a hard problem and heuristics are employed in the D-Wave's embedding function.
Thus, with each run, the number of qubits and the longest chain length (i.e. the number of physical qubits encoding a single qubit) might vary, and even fail at times.
The embedding process can be separately tested.

\begin{lstlisting}[language=Python] 
# Embedding a QUBO model in Chimera graph
import networkx as nx
import dimod
import minorminer
import dwave_networkx as dnx
def embed_qubo_chimera(Q_matrix, plotIt = False):
	connectivity_structure = dnx.chimera_graph(3,3) # try to minimize
	G = nx.from_numpy_matrix(Q_matrix)
	max_chain_length = 0
	while(max_chain_length == 0):
		embedded_graph = minorminer.find_embedding(G.edges(), connectivity_structure)
		for _, chain in embedded_graph.items():
		    if len(chain) > max_chain_length:
		        max_chain_length = len(chain)
	print("max_chain_length",max_chain_length) # try to minimize
	dnx.draw_chimera_embedding(connectivity_structure, embedded_graph)
	plt.show()
\end{lstlisting}

\begin{figure}[ht]
\centering
\includegraphics[width=0.4\textwidth]{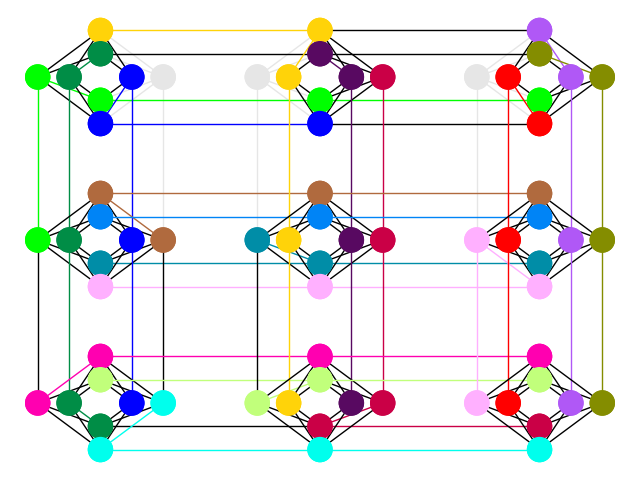}
\caption{Embedding the QUBO in Chimera graph topology for D-Wave Quantum Annealer. Each color represents one of the 16 logical qubits}
\label{f_embed}
\end{figure}

After multiple attempts, the best embedding obtained for our example de novo problem uses 60 qubits with a maximum chain length of 5, as shown in figure~\ref{f_embed}.
Each color represents one of the 16 logical qubits.
 Embedding of our QUBO model in the Chimera graph on the D-Wave Quantum Annealer. (Maybe also mention what the colors mean!)
The following code connects to the D-Wave cloud and solves the de novo example.

\begin{lstlisting}[language=Python] 
# Solve an Ising model using D-Wave solver
from dwave.cloud import Client
from dwave.embedding import embed_ising, unembed_sampleset
from dwave.embedding.utils import edgelist_to_adjacency
from dwave.system.samplers import DWaveSampler
from dwave.embedding.chain_breaks import majority_vote
def solve_ising_dwave(hii,Jij):
	config_file='/media/sf_QWorld/QWorld/QA_DeNovoAsb/dwcloud.conf'
	client = Client.from_config(config_file, profile='aritra')
	solver = client.get_solver() # Available QPUs: DW_2000Q_2_1 (2038 qubits), DW_2000Q_5 (2030 qubits)
	dwsampler = DWaveSampler(config_file=config_file)

	edgelist = solver.edges
	adjdict = edgelist_to_adjacency(edgelist)
	embed = minorminer.find_embedding(Jij.keys(),edgelist)
	[h_qpu, j_qpu] = embed_ising(hii, Jij, embed, adjdict)

	response_qpt = dwsampler.sample_ising(h_qpu, j_qpu, num_reads=solver.max_num_reads())
	client.close()

	bqm = dimod.BinaryQuadraticModel.from_ising(hii, Jij)
	unembedded = unembed_sampleset(response_qpt, embed, bqm, chain_break_method=majority_vote)
	print("Maximum Sampled Configurations from D-Wave\t===>")
	solnsMaxSample = sorted(unembedded.record,key=lambda x: -x[2])
	for i in range(0,10):
		print(solnsMaxSample[i])
	print("Minimum Energy Configurations from D-Wave\t===>")
	solnsMinEnergy = sorted(unembedded.record,key=lambda x: +x[1])
	for i in range(0,10):
		print(solnsMinEnergy[i])
\end{lstlisting}

The top 10 (out of 65536) maximum sampled configurations are shown below.
It is important to note that, the highest sampled configuration is not the global minima in terms of energy, showing the heuristic nature of the annealer.
\begin{lstlisting}[language=bash,numbers=none]
Maximum Sampled Configurations from D-Wave	===>
([-1, -1, -1,  1, -1, -1, -1, -1,  1, -1, -1, -1, -1, -1,  1, -1], -27.9228825, 4562)
([-1, -1,  1,  1, -1, -1, -1, -1,  1, -1, -1, -1,  1, -1, -1, -1], -22.70124548, 611)
([-1, -1,  1,  1, -1, -1, -1, -1, -1,  1, -1, -1,  1, -1, -1, -1], -26.16476862, 481)
([-1, -1,  1,  1, -1, -1, -1, -1,  1, -1, -1, -1,  1, -1, -1, -1], -22.70124548, 474)
([-1, -1, -1,  1, -1, -1, -1, -1, -1,  1, -1, -1, -1, -1,  1, -1], -28.08099638, 470)
([-1, -1, -1,  1, -1, -1, -1, -1,  1, -1, -1, -1, -1, -1,  1, -1], -27.9228825, 343)
([ 1, -1, -1, -1, -1, -1, -1, -1, -1,  1, -1, -1, -1, -1,  1, -1], -27.81747324, 295)
([-1, -1, -1,  1, -1, -1, -1, -1,  1, -1, -1, -1, -1, -1,  1, -1], -27.9228825, 259)
([-1, -1, -1,  1, -1, -1, -1, -1, -1, -1,  1, -1,  1, -1, -1, -1], -27.60665473, 200)
([-1, -1, -1,  1, -1, -1, -1, -1, -1, -1,  1, -1,  1, -1, -1, -1], -27.60665473, 187)
\end{lstlisting}

The top 10 (out of 65536) minimum energy configurations are shown below.
The list shows that, though the D-Wave was able to sample two of the four correct solutions, it has not sampled it with a high probability.
Also, we find two other solution configurations are missed.
Each run of the sampler would be slightly different varying both on environmental errors of the physical qubit system as well as the heuristics of embedding and schedule.
Thus, while we were able to find an acceptable solution by the physical system, it might not be practical for larger problems. 
\begin{lstlisting}[language=bash,numbers=none]
Minimum Energy Configurations from D-Wave	===>
([-1, -1,  1, -1, -1, -1, -1,  1,  1, -1, -1, -1, -1,  1, -1, -1], -30.41886117, 26)
([-1, -1,  1, -1, -1, -1, -1,  1,  1, -1, -1, -1, -1,  1, -1, -1], -30.41886117, 2)
([-1, -1,  1, -1, -1, -1, -1,  1,  1, -1, -1, -1, -1,  1, -1, -1], -30.41886117, 2)
([-1, -1, -1,  1,  1, -1, -1, -1, -1,  1, -1, -1, -1, -1,  1, -1], -30.41886117, 29)
([-1, -1, -1,  1,  1, -1, -1, -1, -1,  1, -1, -1, -1, -1,  1, -1], -30.41886117, 1)
([-1, -1, -1,  1,  1, -1, -1, -1, -1,  1, -1, -1, -1, -1,  1, -1], -30.41886117, 7)
([-1, -1,  1, -1, -1, -1, -1,  1,  1, -1, -1, -1, -1,  1, -1, -1], -30.41886117, 1)
([-1, -1, -1,  1, -1,  1, -1, -1,  1, -1, -1, -1, -1, -1,  1, -1], -29.89181489, 23)
([ 1, -1, -1, -1, -1, -1,  1, -1, -1, -1, -1,  1, -1,  1, -1, -1], -29.89181489, 5)
([-1, -1, -1,  1, -1,  1, -1, -1,  1, -1, -1, -1, -1, -1,  1, -1], -29.89181489, 3)
\end{lstlisting}

A 16 qubit system was required for solving the above problem.
When mapping the 16 qubits to a realistic hardware like D-Wave 2000Q, the connectivity of the qubits in the physical topology is important.
The embedding process considerably increases the number of required qubits and also the quality of the solution.
The highest number of DNA reads that can be solved on a D-Wave 2000Q machine is 9.
The amount of qubits needed to solve the problem grows as $N^2$ and finding embedding for the case with 10 reads will fail in most (if not all) cases.
In classical computation however, the record for exact solutions to the problem, using branch and bound algorithms is 85900 cities TSP~\cite{cook2011pursuit}.
Heuristics like Monte Carlo methods are used for larger inputs.
This experiment infers the need for a much enhanced D-Wave system to do practical de novo assembly.
Steps in this direction can be either in reducing the errors, having a custom anneal schedule, having more qubits and better connectivity (like the Pegasus architecture of the upcoming 5000 qubit model).
At the current state of development, we were able to show a simple proof of concept both on the simulator and the quantum annealer on the cloud.
De novo sequencing on quantum annealers needs to be evaluated with each release of improved hardware to reach a quantum advantage in computation over existing high-performance computing systems. 
\subsection{QUBO using QAOA}

Formulating a problem on QAOA involves specifying the ansatz for the cost and driver Hamiltonian.
Other optimization hyper-parameters involve the initialization circuit, approximation order (cycles), initial parameters and threshold on classical optimizer iterations/precision.
The pseudo-code for the formulation steps in an OpenQL implementation is shown in Listing~\ref{code:pqc_use}.


First, the TSP city-graph is created based on the example in section~\ref{s5p1}.
The networkx Python package is used to create a directed weighted complete graph based on the pair-wise read edge-weights.

The graph is then converted to the problem Hamiltonian.
The problem Hamiltonian is stored as a weighted sum-of-product of Paulis.
For $n$ cities (TSP graph nodes), $n^2$ qubits are required representing the city nodes and the time slots, encoded as:
$$q0\dots q15 \equiv [n_0t_0 | n_0t_1 | n_0t_2 | n_0t_3 | n_1t_0 | n_1t_1 | n_1t_2 | n_1t_3 | n_2t_0 | n_2t_1 | n_2t_2 | n_2t_3 | n_3t_0 | n_3t_1 | n_3t_2 | n_3t_3]$$
On each qubit, there can be either of the 4 Pauli operators, $\{I,X,Y,Z\}$, thus a maximum of $4^{n^2}$ weighted sum-of-product Pauli terms are possible.
This amounts to $4294967296$ Pauli terms when $n=4$ (in our example), thus, we store only the non-zero terms.
For TSP optimization, however, only the $\{I,Z\}$ operator is required.

Firstly, each city and each time-slot must be assigned, but not all together.
Thus, a term is added with a positive penalty ($w = 100000.0$) for each qubit (the term being a $Z$ operator on the specific qubit and $I$ otherwise).
We will abbreviate the sum-of-product of Pauli term notation henceforth by assuming Identity for qubits not mentioned in a Pauli term.
Thus: $$w*\{ZIIIIIIIIIIIIIII+IZIIIIIIIIIIIIII+IIZIIIIIIIIIIIII+\dots+IIIIIIIIIIIIIIIZ\}$$
$$ H_C^1 = w*\{Z_0+Z_1+Z_2+\dots+Z_{15}\} = \sum_{q=0}^{n^2-1} wZ_q$$

Then, for each penalty of co-location (two cities, same time slot), a term with 2 $Z$ operators for the two conflicting qubits is added with a positive penalty weight, and two separate terms for each penalty qubits with a 1 $Z$ are added with a negative penalty.
$$H_C^2 = \sum_{r=0}^{n-1} \sum_{i=1}^{n-1} \sum_{j=0}^{i-1} \Bigg\{-\dfrac{w}{2}Z_{in+r} -\dfrac{w}{2}Z_{jn+r} +\dfrac{w}{2}Z_{in+r}Z_{jn+r}\Bigg\}$$

Similar terms are added for repetition (two time slots, same city).
$$H_C^3 = \sum_{i=0}^{n-1} \sum_{r=1}^{n-1} \sum_{s=0}^{r-1} \Bigg\{-\dfrac{w}{2}Z_{in+r} -\dfrac{w}{2}Z_{in+s} +\dfrac{w}{2}Z_{in+r}Z_{in+s}\Bigg\}$$

For each TSP edge, the edge pair is assigned to consecutive time slots.
The edge weight is added as a penalty for the 2 $Z$ operator terms, while assigning the individual terms with a negative penalty.
$$H_C^4 = \sum_{i=0}^{n-1} \sum_{\substack{j=0\\j\neq i}}^{n-1} \sum_{\substack{r=0\\s=(r+1)\%n}}^{n-1} \Bigg\{ -\dfrac{d_{ij}}{4}Z_{in+r} -\dfrac{d_{ij}}{4}Z_{jn+s} +\dfrac{d_{ij}}{4}Z_{in+r}Z_{jn+s}\Bigg\}$$
Thus, these negative penalty terms in the overall equation stand out among the positive penalty terms for all qubits in $H_C^1$, so that only a valid path is assigned as a solution.
The final cost Hamiltonian is $H_C = H_C^1+H_C^2+H_C^3+H_C^4$

Now the ansatz needs to be formed.
For the cost Hamiltonian ansatz, for each single $Z$ terms, the Pauli-Z is replaced with a parameterized Z-Rotation gate $R_Z(\gamma)$.
The double $Z$ terms between qubits $(a,b)$ are replaced with a $CNOT_{ab}R_Z^b(\gamma)CNOT_{ab}$ (where $b$ is the target qubit).
For the mixing Hamiltonian, a $R_X(\beta)$ is used on all qubit (or $HR_Z(\beta)H$).

\begin{lstlisting}[numbers=none,mathescape=true,columns=fullflexible,backgroundcolor=\color{light-gray},mathescape=true,frame=lines,caption=Pseudo-code for applying QAOA in OpenQL for optimization,label={code:pqc_use}]
$\bullet$ PQC encoding (e.g. Travelling salesman problem, Maximum cut problem):
    $\diamond$ Create (weighted) graph with networkx
    $\diamond$ Converts graph to weighted-Sum-of-Product-of-Paulis (wsopp). Encoding depends on problem. This is the problem/cost Hamiltonian for QUBO/Ising model.
    $\diamond$ Convert graph to ansatz with cost and mixing Hamiltonian as parameterized QASM (ansatz, coefficients, angle ids)

$\bullet$ Initialization:
    $\diamond$ Classical optimizer object from SciPy:
        $\star$ name [Default: Nelder-Mead]
        $\star$ convergence function tolerance [Default: 1.0e-6]
        $\star$ iteration limit
    $\diamond$ Reference/initial state quantum circuit [Default: equal superposition, Hadamard on all qubits]
    $\diamond$ Steps (ansatz blocks per iteration) [Default: 1]
    $\diamond$ Parameters:
        $\star$ for cost Hamiltonians (gammas) for each step
        $\star$ for mixing/driving Hamiltonians (betas) for each step
      [Default: random angles in 0 to 2$\pi$]
    $\diamond$ Shots (for state tomography measurement aggregate) [Default: 0, QX internal state vector is accessed]

$\bullet$ Invoke QAOA
\end{lstlisting}

The reference state is set to an equal superposition (Hadamard over all qubits).
The cost and mixing Hamiltonian ansatz is alternated for the set number of steps (1 in our case).
This forms the parametric circuit for the quantum computer.
Along with this, the random initial parameters vector $(\Gamma,B)$ is passed to the classical optimizer wrapper for the QAOA.
The QAOA is run as explained in listing~\ref{code:pqc}.

\begin{figure}[ht]
\centering
\includegraphics[clip, trim=10cm 0cm 0cm 8cm,width=0.9\textwidth]{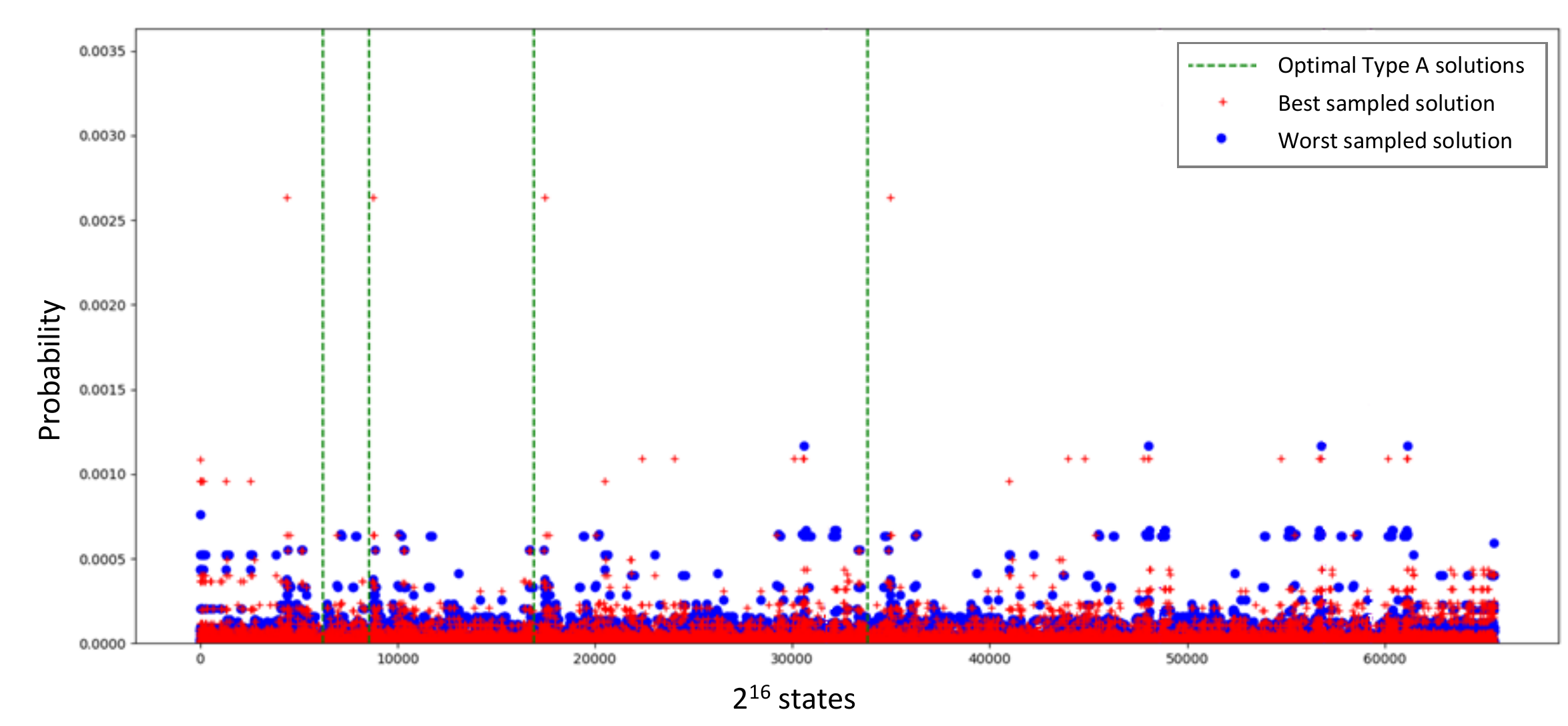}
\caption{Result of simulating QuASeR example using QAOA on the QX Simulator}
\label{f_resqaoa}
\end{figure}

We simulated our algorithm on the QX Simulator to validate the results. 
To speed up the simulation, we access the internal state vector from QX Simulator instead of applying the state tomographic trials.
However, optimizing the parameters on 16 qubits for single iteration over a single QAOA step proved to be cumbersome, 
reaching the limit of work memory for most runs.
While we obtain good results for the Max-Cut problem (which requires as many qubits as nodes in the graph), the $n^2$ qubit space for TSP is costly for simulation.
Similarly, most online tools only offer examples for the trivial case of an undirected triangle graph (where only one Hamiltonian cycle is possible).
We used the Nelder-Mead, Powell and BFGS optimizers in SciPy.
We found that the optimizers are able to explore only a small space near the initial guess before settling at a suboptimal solution.
This is shown in figure~\ref{f_resqaoa}, where the 4 optimal type A solution states out of the $2^{16} = 65536$ basis states are indicated by the dotted vertical green lines.
The experiment uses a QAOA depth of 1, a random angle initialization, an equal superposed initial state (Hadamard on all qubits) and 40 reruns for the QAOA optimizer.
When the initial guess is bad, the highest probability states (high blue circles) are not close to the optimal lines, whereas for good initial conditions case (red pluses), the optimizer eventually reach quite close to the optimal.
The optimal solutions lie in near vicinity of the found solutions (high red pluses) and can be explored via exhaustive search near these sub-optima suggested by QAOA.
The dotted line (in blue) represents the initial guess while the solid line (in orange) represents the final solution.
QAOA is able to improve 2 out of the 4 solutions of Type A. 
Development of new classical optimizers~\cite{guerreschi2017practical,khairy2019learning} and their hyper-parameter settings~\cite{shaydulin2019multistart} for HQC algorithms is a research field of its own, which is not the focus of this paper.
Thus, while we were able to formulate a generic de novo assembly problem to QAOA, we were not able to obtain satisfactory results from the simulation.
This motivates the need for both faster simulation and the access to better NISQ devices where this entire pipeline can be executed, a common challenge for quantum computing today.

An approximation algorithm solves every instance of a given NP-hard problem with some guaranteed quality in expectation. 
The value of merit is the ratio between the quality of the polynomial-time solution and the quality of the true solution.
The reason for the interest in QAOA is its potential to exhibit quantum supremacy.
QAOA, though promising for exhibiting "quantum supremacy" does not imply that it will be able to outperform classical algorithms on important combinatorial optimization problems such as Constraint Satisfaction Problems.
Current implementations of QAOA are subject to a gate fidelity limitation, where the potential advantages of larger values of the parameter $p$ in QAOA applications are likely to be countered by a decrease in solution accuracy. 
\section{Conclusion} \label{s7}

In this paper, we formulated a de novo assembly sequence reconstruction algorithm called QuASeR using the overlap-layout-consensus approach for acceleration on a quantum computing platform.
The quantum kernel is formulated for both a gate-based quantum system as well as a quantum annealer.
The required technical background for formulating the de novo sequencing problem (i.e. QUBO, TSP, and Hamiltonians) is introduced with simple examples to target both the genomics research community and quantum application developers.
A proof-of-concept de novo sequence assembly is mapped to TSP and then to a QUBO.
This is firstly solved on the D-Wave simulator and D-Wave Quantum Annealer.
All 4 correct results are obtained on the simulator while only 2 of the solutions are sampled on the Quantum Annealer (though with less probability).
The connectivity topology of the D-Wave architecture limits embedding larger problem instances.
The variational algorithm approach for gate-based quantum computing is introduced for solving optimization problems using QAOA. 
This algorithm performs two optimization steps, one executed on a quantum circuit and another on a classical computer.
The proposed de novo algorithm is solved using QAOA, and then simulated on the QX simulator.
Simulation showed that the results are heavily dependent on the exploratory capabilities of the classical optimizer.
A gate-based quantum computer is not targeted as the coherance time, connectivity topology and number of qubits prevents any meaningful result at the current state of available quantum hardware.
This research is part of a full-stack domain-specific quantum accelerator project undertaken in the Quantum Computer Architecture lab at the Delft University of Technology.
It is the first time this important computation problem in bioinformatics is attempted on a quantum accelerator.

\bibliographystyle{unsrt}
\bibliography{ref}

\end{document}